**Epistasis and the structure of fitness landscapes: are experimental fitness landscapes compatible with Fisher's geometric model?**

François Blanquart[1,2*], Thomas Bataillon[1].

1. Bioinformatics Research Centre, Aarhus University. 8000C Aarhus, Denmark.

2. Department of Infectious Disease Epidemiology, Imperial College London, St Mary's Campus, London, United Kingdom.

François Blanquart

Department of Infectious Disease Epidemiology

Imperial College London, St Mary's Campus

Norfolk Place, W2 1PG London

United Kingdom.

francois.blanquart@normalesup.org

Key words: fitness landscape, mutational network, epistasis, adaptation, Fisher's geometric model, antibiotic resistance

**Abstract**

The fitness landscape defines the relationship between genotypes and fitness in a given environment, and underlies fundamental quantities such as the distribution of selection coefficient, or the magnitude and type of epistasis. A better understanding of variation of landscape structure across species and environments is thus necessary to understand and predict how populations will adapt. An increasing number of experiments investigates the properties of fitness landscapes by identifying mutations, constructing genotypes with combinations of these mutations, and measuring the fitness of these genotypes. Yet these empirical landscapes represent a very small sample of the vast space of all possible genotypes, and this sample is often biased by the protocol used to identify mutations. Here we develop a rigorous statistical framework based on Approximate Bayesian Computation to address these concerns, and use this flexible framework to fit a broad class of phenotypic fitness models (including Fisher's model) to 26 empirical landscapes representing 9 diverse biological systems. In spite of uncertainty due to the small size of most published empirical landscapes, the inferred landscapes have similar structure in similar biological systems. Surprisingly, goodness of fit tests reveal that this class of phenotypic models, which has been successful so far in interpreting experimental data, is a plausible model in only 3 out of 9 biological systems. More precisely, although Fisher's model was able to explain several statistical properties of the landscapes – including mean and standard deviation of selection and epistasis coefficients –, it was often unable to explain the full structure of fitness landscapes.

**Introduction**

The fitness landscape is defined by a set of genotypes, the mutational distance between them and their associated fitness in a given environment (Wright 1931; Orr 2005). The structure of the fitness landscape determines the fitness effects of mutations, and the interaction between mutations for fitness. These properties determine the pace of adaptation (Eyre-Walker & Keightley 2007), the predictability of evolution (Weinreich et al. 2006), the benefits of sexual reproduction (Kondrashov & Kondrashov 2001; de Visser et al. 2009), and the probability of speciation (Gavrilets 2004; Chevin et al. 2014). Thus, it is an important goal of evolutionary biology to characterize experimentally the properties of fitness landscapes across species and environments (de Visser & Krug 2014).

The most straightforward and popular experimental approach to access the properties of the fitness landscape consists in identifying mutations, constructing several genotypes that only differ by various combinations of these mutations, and measuring the fitness of these genotypes. This protocol allows reconstructing what we call "empirical landscapes". For example, several experiments identify a small number, $L$, of mutations and consider the fitness of $2^L$ genotypes with all possible combinations of these mutations. Early studies were primarily descriptive, with a focus on patterns of epistasis among mutations (Malcolm et al. 1990; de Visser et al. 1997; Whitlock & Bourguet 2000). In an influential study, Weinreich et al. (Weinreich et al. 2006) studied the landscape between an ancestral strain of *Escherichia coli* and an evolved type with 5 mutations conferring high antibiotic resistance. They computed the number of paths up to the fitness maximum that could be followed by population evolving by natural selection, and showed that the ruggedness of the landscape implied that very few mutational paths could be used during biological evolution. This study suggested that the structure of fitness landscapes might severely constrain evolutionary trajectories, thus opening up the possibility that adaptation could be predicted to some extent. This finding has inspired the characterization of many other empirical landscapes (reviewed in (Weinreich et al. 2013)).

In principle, empirical landscapes can be compared with predictions from theoretical fitness landscape models. For example, several studies fit specific models to empirical landscapes (Lunzer et al. 2005; Chou et al. 2011; Rokyta et al. 2011; Schenk et al. 2013; Chou et al. 2014). These models predict quantitatively the fitness values and epistasis coefficients, and as such greatly improve our understanding of the form of epistasis that is typical of the particular system under study. However, the increasing number of empirical landscapes calls for a more general method to infer and compare the properties of fitness landscapes across species and environments. This possibility is very appealing, and timely given that data accumulates on a diversity of empirical systems and selective environments, but raises several challenges.

The variability observed between empirical landscapes might be driven by biological differences of interest between organisms and environments of selection, but this variability is currently confounded with two other factors: stochastic variability due to sampling of a small number of mutations, and variability in the protocol by which mutations are isolated. The full fitness landscape of a species in the environment of selection is defined as the fitness of all possible genotypes in this environment. This is an incredibly large space, scaling exponentially with the size of the genome. Most experiments explore a very small subset of the landscape, as they examine at best a few dozens of genotypes. Starting from the ancestral genotype, a single point in this large fitness landscape, the region of the fitness landscape that is explored depends on the particular mutations that were isolated. Thus, each empirical landscape results from a single realization of the stochastic sampling of a small number of mutations from a myriad of available mutations (Tenaillon et al. 2007; Salverda et al. 2010; Schenk et al. 2013; Szendro et al. 2013; de Visser & Krug 2014). In other words, a single constant underlying fitness landscapes can give rise to a diversity of small genotypic landscapes depending on the mutations that are sampled (Blanquart et al. 2014). Moreover, the region of the underlying fitness landscapes that is explored depends on the experimental protocol used to isolate mutations. For example, mutations are often obtained under protocols involving natural selection. While random mutations give more rugged empirical landscapes, mutations that have been sequentially selected in a single population give smoother empirical landscapes (Szendro et al. 2013; Draghi & Plotkin 2013; Blanquart et al. 2014). Thus, inferring the properties of fitness landscapes from empirical data in meaningful ways requires (i) quantifying the uncertainty due to sampling of a limited number of mutations, and (ii) explicitly modelling how mutations were experimentally isolated.

In this study, we address these challenges and develop a statistical framework to infer the properties of the underlying fitness landscape from empirical landscapes. We use a broad class of phenotypic fitness landscape models that includes Fisher's geometric model (Fisher 2000). Phenotypic fitness landscapes model how the genotype of an organism translates into a set of phenotypes, which themselves determine fitness. In other words, the very large space of all possible genotypes is projected onto a continuous phenotypic space of arbitrary dimensionality, and fitness depends only on the position in this phenotypic space. Fisher's model, in particular, assumes that the phenotypes are under stabilising selection towards a single optimum, that the effects of mutations in the phenotypic space are drawn from a multivariate Gaussian distribution, and that mutations combine additively in the phenotypic space. Phenotypes can be biological traits that need to be tuned to a precise level to maximize growth of the organism in the environment of selection, for example the concentration of an enzyme in a metabolic pathway, or the level of expression of a gene. Fisher's model can also be viewed as an abstract statistical description of the genotype-fitness map.

A number of reasons motivate the choice of Fisher's model as underlying fitness landscape. A phenotypic model solves the problem of high dimensionality of the genotypic space. Indeed genotypic fitness landscape models such as the Rough Mount Fuji model (Szendro et al. 2013) or the NK model (Kauffman & Levin 1987) require a number of parameters increasing linearly with the number of mutations or the number of genotypes. In contrast a phenotypic model can describe an arbitrary large number of genotypes using a small number of parameters. More fundamentally, it has recently been shown that Fisher's model emerges from a set of "first principles" that specifies how fitness results from developmental integration of a large number of mutable traits (Martin 2014). Last, Fisher's geometric model is simple yet can generate a diversity of empirical landscapes (Blanquart et al. 2014), and it successfully predicts experimental quantities, such as the distribution of epistasis coefficient between pairs of mutations (Martin & Lenormand 2006; Martin et al. 2007), and the dynamics of mean fitness over time (Perfeito et al. 2014).

This study focuses on the following questions: How much information on the structure of the underlying fitness landscape can be inferred from existing empirical landscapes? What are the properties of fitness landscapes inferred from empirical data available so far, and are underlying landscape similar in similar species or environments? Is the structure of empirical landscapes compatible with a model assuming stabilizing selection on a set of underlying, unknown phenotypes?

To answer these questions we develop an inference framework that allows fitting Fisher's model to a diversity of experimental datasets obtained under a range of protocols. Using this framework, we infer the parameters and quantify the goodness of fit of Fisher's model on 26 published genotypic landscapes representing 9 distinct biological systems. We infer the properties of the underlying fitness landscape of each dataset while accounting for the protocol used to obtain data, allowing a meaningful comparison of fitness landscapes inferred across several species and environments. This survey reveals substantial differences in the shape of underlying fitness landscapes across biological systems and environments of selection. We also show that Fisher's model is able to fully account for the observed properties of genotypic landscapes in only 3 biological systems out of 9.

**Materials and Methods**

Dataset selection

We searched the literature for published empirical landscapes that include clearly identified sets of genotypes with combinations of two mutations or more, together with their fitness. The way in which these mutations evolved or were obtained had to be sufficiently described so we could reproduce it with simulations (see below). For selected mutations, we verified that the fitness measure reported is relevant to the environment in which the mutations evolved. We identified a total of 26 published datasets spanning 9 independent biological systems meeting these criteria. In the following we will identify the datasets representing these nine systems using the letters A to I (Table 1, and Supplementary Information). The datasets encompassed a diversity of species including species of virus, bacteria, fungi, animals, and of ecological scenarios (Table 1). Several experiments explored the fitness landscape of species in a lab environment using random mutations, in the fungus *Aspergillus niger* (de Visser et al. 1997) (A1 and A2), the fly *Drosophila melanogaster* (C1-C2) (Whitlock & Bourguet 2000), and the budding yeast *Saccharomyces cerevisiae* (Costanzo et al. 2010). The latter dataset is a large collection of 5596 deletion mutants. To reduce this large dataset to a size amenable to our analysis, we randomly drew 10 independent, randomly chosen subsets that included 20 mutations, all single mutants and 100 double mutants (all combinations of the first 10 mutations times the last 10 mutations, for a total of 121 possible genotypes, but in reality 104 to 116, as some genotypes were missing).

Three datasets represented the fitness landscape of two virus species adapting to their hosts (D, E1, E2) (Rokyta et al. 2011; Sanjuán et al. 2004). Two datasets represented landscapes of adaptation of microbial species to a novel environment, including a long-term selection experiment in low-glucose environment (F) (Khan et al. 2011) and a selection experiment in methanol environment (G) (Chou et al. 2011). Last, seven datasets represented empirical landscapes reconstructed from mutations that confer drug resistance. These included studies of mutations in the enzyme TEM-1 beta-lactamase, which confer resistance to cefotaxime resistance in bacteria (four datasets H1-H4) (Weinreich et al. 2006; Tan et al. 2011; Schenk et al. 2013) and studies of mutations in the dihydrofolate reductase gene, which confer pyrimethamine resistance (an antimalarial drug) in transgenic bacteria and yeast (three datasets I1-I3, (Lozovsky et al. 2009; Brown et al. 2010; Jiang et al. 2013)).

Data analysis

A variety of fitness measures were reported in the published empirical landscapes we collected. Our analysis requires meaningful estimates of fitness value to model how selected mutations differ from random mutations.

Meaningful selection coefficients are expressed in units of log-fitness. They must be calculated either as $\log[\lambda_m/\lambda_0]$, where $\lambda_m$ and $\lambda_0$ are the multiplicative growth rate of the mutant and the ancestor (called "fitness" in most population genetics model), or as $r_m - r_0$, where $r_m$ and $r_0$ are exponential growth rates (Chevin 2011). Unfortunately, many studies only reported the ratio $r_m/r_0$ (Table 1, landscapes A, B, E, F, G, I3), which in theory cannot be used to obtain a correct selection coefficient. To analyse the studies that only report $r_m/r_0$, we used $\log[r_m/r_0]$ as a log-fitness measure. This measure is approximately equal to $(r_m - r_0)/r_0$ under weak selection, which is a quantity proportional to the selection coefficient. Moreover, this log-fitness measure, conveniently, does not depend on the unit of the growth rate and can be compared across landscapes.

For drug resistance fitness landscapes, only one dataset reported growth rate at a given drug concentration (landscape I3, Table 1). Other studies reported the Minimum Inhibitor

Concentration (MIC) or a similar measure (Table 1, landscapes H1-H2 and I1-I2). MIC, the concentration of drug above which the population cannot grow, is not easily related to fitness. For this reason we presented the results of MIC landscapes in Supplementary Information (Weinreich et al. 2006; Tan et al. 2011; Schenk et al. 2013).

We proceeded to several additional steps of data cleaning. Non-viable genotypes (fitness value of 0) were excluded from the analysis (three genotypes in total, one in a pyrimethamine landscape I1, two in a Drosophila landscape C2), because Fisher's model cannot easily account for lethal mutations. In dataset G, the order of fixation of co-selected mutations was unknown. We assumed mutations fixed from the largest effect mutation to the smallest effect mutation, in accordance with the reported dynamics of mean fitness through time in the experiment. In dataset I2, two mutations occurred at the same locus. We made this dataset compatible with our framework (which assumes each locus is diallelic) by excluding all genotypes bearing the third allele.

Approximate Bayesian Computation

Table 1 shows that a variety of protocols was used to obtain empirical landscapes. Some of the empirical landscapes were formed of single and double mutants only, while other included all possible combinations of 4 or 5 mutations, thus including genotypes with 3, 4 or 5 mutations. Moreover, the way in which mutations were isolated also varied. Mutations were random, independently selected, or co-selected. "Independently selected" means that mutations emerged under the action of selection in separate populations evolving independently from a unique ancestral genotype. "Co-selected" means that mutations were selected sequentially in the same population. Modelling the way selection biased the resulting empirical landscape is already complicated. To make matters worse, several protocols included an additional step. These protocols were used to study the landscape of resistance to cefotaxime, a beta-lactam antibiotic (landscapes H1-H4). Among a large set of 48 mutations found individually in cefotaxime-resistant natural isolates, three smaller subsets were studied in details. These subsets were composed of the four mutations of smallest fitness effect, the four mutations of largest fitness effect (H3-H4), and five mutations that together conferred a very high fitness (H1-H2). To account for this variety of protocols, we used a flexible Approximate Bayesian Computation (hereafter ABC) approach to infer from empirical data the parameters underlying Fisher's geometric model.

(i) Details of the Approximate Bayesian Computation framework

The original ABC "rejection algorithm" proceeds as follows. A large number of parameter sets is drawn in a prior distribution. For each parameter set $\theta$, a dataset $\widehat{D}(\theta)$ is simulated and a measure of distance between the true dataset and each simulation $\rho\big(\widehat{D}(\theta), D\big)$ is computed. A set of parameters is retained in the posterior distribution if the distance between $D$ and $\widehat{D}(\theta)$ is lower than a small value $\epsilon$. In other words, the posterior distribution is composed of all the parameter sets $\theta$ such that $\rho\big(\widehat{D}(\theta), D\big) < \epsilon$. In practice, $\epsilon$ is chosen such that a given, small fraction of the prior parameter sets is retained in the posterior (Csilléry et al. 2012), but ABC will give the correct posterior distribution of parameters only in the limit where $\epsilon$ is close to zero.

The distance between the dataset and simulation is often defined based on a set of statistics. This set of statistics must be carefully chosen to be informative but of relatively low dimensionality. We conducted the analysis using either the full set of observed log-fitness values (16 to 121 fitness values), or a set of 6 summary statistics. The 6 summary statistics are as follows. (i) The mean coefficient of selection of all single mutants. (ii) The mean epistasis coefficient between all pairs of mutations, averaged over all genetic backgrounds. (iii) The standard deviation of selection coefficients. (iv) The standard deviation of epistasis. (v) The correlation between the epistasis coefficient and the background fitness. Specifically, for each pair of mutations we calculate the epistasis coefficient and the average fitness of the two genotypes with one of the mutations, and compute the correlation between these two quantities across all pairs of mutations and all genetic backgrounds. (vi) The maximal fitness value (Table S1). The distance of each simulated dataset to the experimental dataset was:

$$\rho(\widehat{D}, D) = \sqrt{\sum_{i=1}^{n_{stat}} \left(\frac{\hat{S}_i - S_i}{\text{mad}(\hat{S}_i)}\right)^2}$$

where $n_{stat}$ is the number of statistics, $S_i$ is the statistic $i$, $\hat{S}_i$ is the simulated statistic $i$. Statistics are normalized by the median absolute deviation, $\text{mad}(\hat{S}_i)$, which is analogous to standard deviation but with medians instead of means. When statistics were the full set of fitness values, genotypes were uniquely identified by ordering mutations by their fitness effects.

We detailed above the "rejection" algorithm, where the posterior is simply the fraction of parameters randomly drawn from the prior distribution that generates simulated landscapes closest to the data. For this algorithm we used a tolerance (the fraction of retained simulations) of 0.005 (using the lower tolerance of 0.0005 did not improve accuracy). In addition to the rejection algorithm, we used a linear regression algorithm (Beaumont et al. 2002). In this method, the posterior parameters are corrected using a local-linear regression of the parameter values onto the summary statistics, giving more weight to simulations closer to the dataset. Last, we used a "neural network" algorithm that adjusts the posterior distribution based on a non-linear regression using neural networks (Blum & François 2010). The three methods are implemented in the R package "abc"(Csilléry et al. 2012; R Development Core Team 2010).

(ii) Details of the evolutionary simulations

We simulated a large number of genotypic landscapes under Fisher's model, seeding the simulation with parameters $\theta$ drawn from some prior distributions (detailed below).

The simulated landscapes were based on Fisher's model, a phenotypic fitness landscape model whereby an organism is evolving under stabilizing selection on $n$ continuous phenotypic traits that together determine fitness. Each genotype is characterized by a phenotype vector $\mathbf{z} = \{z_1, z_2, ..., z_n\}$ consisting of $n$ traits, where $n$ is the dimensionality of the phenotypic space. The parameter $n$ defines the number of phenotypes under selection, or "complexity", for an organism evolving in a given environment (Tenaillon et al. 2007; Lourenço et al. 2011; Chevin et al. 2014). The effects of mutations are assumed to be additive in the phenotypic space. For example, if we consider five mutations at five distinct loci of the genome, the genotype 00101, where the series of 0 and 1 denote the absence or presence of mutations at each of five loci (relative to an ancestral strain with genotype 00000), has phenotype $\mathbf{z_0} + \mathbf{dz_3} + \mathbf{dz_5}$, where $\mathbf{z_0}$ is the phenotype vector of the ancestral strain, $\mathbf{dz_3}$ and $\mathbf{dz_5}$ are the phenotypic effects at mutations at loci 3 and 5. The effects of mutations on phenotypes (the vectors $\mathbf{dz}$) are drawn from a multivariate normal distribution with mean $\mathbf{0}$ and variance-covariance matrix $\sigma \, \mathbf{I}_n$, where $\sigma$ is the size of mutations. Thus, each mutation affects jointly all phenotypes (assumption of full pleiotropy). The mapping of phenotype on fitness is defined by $\log[W(\mathbf{z})] = \log[W_{max}] - \|\mathbf{z}\|^Q + e$, where $W_{max}$ is the maximal fitness, which determines the distance to the optimum of the ancestral strain, $\|\mathbf{z}\|$ is the Euclidean norm of the phenotype vector, and $e$ is the experimental error on fitness measurements. Following Wilke and Adami (Wilke & Adami 2001) and others (Tenaillon et al. 2007; Gros et al. 2009), we extended Fisher's geometric model with the parameter $Q$, which quantifies how flat the peak is at the optimum (Fig. 1). Fisher's model *sensu stricto* is the special case where $Q = 2$ – i.e, the fitness function is Gaussian. Our definition of fitness implies that the ancestral strain had log-fitness 0, corresponding to the phenotype $\mathbf{z_0} = \{-\log[W_{max}]^{\frac{1}{Q}}, 0, 0, ...\}$. This normalisation was done without loss of generality. Maximum fitness $W_{max}$, which is the height of the fitness peak in the environment where fitness is measured, was achieved when all phenotypes are at their optimal value, chosen here to be $z = 0$ without loss of generality. Lastly, *e* is the measurement error in log-fitness measure and was assumed to be normally distributed with mean 0 and standard deviation estimated from the empirical data (SI). Figure 1 shows several examples of a single empirical genotypic landscape generated by sampling a small number of mutations in the underlying landscape.

For each set of parameters $\theta = (W_{max}, \sigma, n, Q)$, we simulated the process by which mutations were isolated and generated a genotypic landscape. In practice, the sets of genotypes were of two broad categories: either 4-5 mutations were isolated and genotypes bearing all possible combinations of these mutations ($2^4$ or $2^5$) were constructed, or a larger number of mutations (7 to 9) were isolated and single mutants and double mutants were constructed. Mutations were, depending on the empirical protocol used for obtaining the data, considered to be random, independently selected, or co-selected. For random mutations, simulations consisted in drawing the phenotypic effects of mutations in the multivariate normal distribution $(0, \sigma I_n)$ and then combining these mutations additively and computing fitness using our phenotype to fitness mapping. When mutations were isolated in an experiment involving selection, we assumed adaptation proceeded by successive invasion of beneficial mutations, without clonal interference. This allowed us to conduct fast simulations based on the "Strong Selection, Weak Mutation" (SSWM) approximation (Kimura 1983; Gillespie 1991), making it possible to conduct the large number of simulations required by ABC. Under the SSWM regime, a selected mutation is drawn among the pool of random mutations with each mutation weighted by $\max[0, s]$ where $s$ is the fitness effect of the mutation. This derives from the fact that the probability of fixation of a beneficial mutation is scaling linearly with its fitness effect $s$ in this regime (Patwa & Wahl 2008). Fitness effects were calculated relative to the ancestor for independently selected mutations, and relative to the genetic background with previously evolved mutations for co-selected mutations. For the protocol where 5 mutations which together confer a large fitness effect are isolated (Weinreich et al. 2006), we chose the set of 5 mutations that confers the highest fitness among 1000 random combinations.

For each empirical landscape, $10^6$ genotypic landscapes were generated using $10^6$ parameter sets drawn from prior distributions. Priors were chosen to be uninformative and to ensure that they could generate a diversity of fitness landscapes (Fig. 1). The height of the peak in log-fitness, $\log[W_{max}]$, was drawn from an exponential distribution with mean $2$. Maximum fitness on a log scale ranged from $3.7 \times 10^{-7}$ to $29$ (2.5% - 97.5% quantile $0.05 - 7.4$). The complexity of the phenotypic space, the number of phenotypic dimensions under selection, was given by $n = \lfloor \eta \rfloor + 1$ where $\lfloor . \rfloor$ denotes the floor function and $\eta$ was drawn from an exponential distribution with mean $5$. It ranged from 1 to 75 (2.5% - 97.5% quantile 1-7). We used an exponential prior for complexity because, under Fisher's model with full pleiotropy, the distribution of fitness effects has unrealistically small variance at high complexity. The size of mutations $\sigma$ in the phenotypic space was drawn from an exponential distribution with mean $0.2$. It ranged from $1.7 \times 10^{-7}$ to $2.6$ (2.5% - 97.5% quantile 0.005 - 0.74). The choice of an exponential distribution was motivated by the fact that variations in fitness are modest in many of the datasets, and therefore mutational effects are probably small. The shape of the peak $Q$ was drawn from a uniform distribution $[0.5, 4]$ (Fig. 1).

Cross-validation

We checked the accuracy of inference from empirical landscapes using simulated pseudo-datasets generated under Fisher's model. We performed cross-validation using $n_{CV} = 500$ pseudo-datasets generated under Fisher's model, for each type of experimental protocol (Fig. 2, Table 2). We applied the ABC algorithm on each dataset and compare the posterior distribution of parameters to the true (known) parameters. We computed the prediction error, defined for each parameter as $\frac{\sum(\tilde{\theta}_i - \theta_i)^2}{n_{CV} \, \text{V}[\theta]}$ where $\theta_i$ is the true value of parameter used for the $i$th simulated pseudo-data, $\widetilde{\theta}_\iota$ is the median of the posterior distribution, and $V[\theta]$ is the variance of the prior distribution. The expected prediction error is 0 when inference is perfect (the median always matches the true parameter), and is 1 when no inference can be made (the posterior parameters are drawn at random from the prior). For cross-validation we assumed experimental errors were 0 in order to compare the accuracy of inference across protocols in an ideal case where fitness values are perfectly known.

Posterior predictive checks

We next tested whether the empirical landscapes we analysed were compatible with the hypothesis that Fisher's landscape was the true model for the empirical data. We used posterior predictive checking (Gelman et al. 2014) to quantify the goodness of fit of Fisher's model to each dataset. For each experimental dataset, we ran the ABC algorithm on 1000 random pseudo-datasets generated using parameters drawn from the joint posterior distribution of parameters. For each of these pseudo-datasets, we recorded the median distance between the pseudo-dataset and the accepted (closest) simulated data in the ABC algorithm. This resulted in a null distribution for the median distance of the simulations retained in the ABC algorithm, which is the distribution of distance between simulations and data when Fisher's model is truly underlying the data. We then used this distribution to compute a "Bayesian p-value", also known as posterior predictive p-value in Bayesian model checking (Gelman et al. 2014). This p-value is the probability that median distances for pseudo-datasets generated under Fisher's model are greater than the median distance of the experimental dataset. A low p-value suggests that the data is further apart from Fisher's model simulations than expected if the data followed Fisher's model. A p-value was computed for the distances based on summary statistics and for the distances based on all fitness values. For the latter, we also decomposed the distance and computed an analogous p-value for each individual genotype, to identify genotypes with fitness values that are particularly unlikely under Fisher's model (those whose individual p-value is lower than 0.05).

## Results

*Cross-validation and accuracy of parameter inference*

We quantified the accuracy of inference from empirical landscapes using 500 simulated pseudo-datasets generated under Fisher's model. This analysis revealed that the true parameters of the underlying landscape are generally inferred with mediocre accuracy under most protocols used in existing studies (Fig. 2, Table 2). Inference based on summary statistics (Table 2) always yielded lower error than inference based on all fitness values (Table S2). Using summary statistics makes the ABC algorithm more accurate because it alleviates the "curse of dimensionality": the distance of the data to the accepted simulations is closer to 0 for the same number of simulations, such that the main assumption of ABC is better respected. However, the use of summary statistics causes loss of information (Sünnaker et al. 2013). Here, the gain of accuracy more than offset the loss of information, making inference based on summary statistics better.

ABC is an approximate method, and we cannot rule out totally that low accuracy was due to these approximations. However, low accuracy may also be caused by the limited information contained in small genotypic landscapes. In other words, even if the inference method was perfect, the true posterior distribution of parameters may still be quite wide and cause low accuracy. Because we have explored a number of variations on the ABC algorithm, including three different algorithms, full statistics versus summary statistics, and several values of tolerance (Fig. S1), and accuracy of inference was always relatively low, we hypothesise that the main reason behind low accuracy was probably the limited information contained in genotypic landscapes. Each empirical landscape conveys rather modest information on the underlying, "true" fitness landscape.

In particular, empirical landscapes conveyed almost no information on the number of phenotypes under selection ($n$). Prediction errors for this parameter were always higher than 0.5 and often close to 1. The size of mutations $\sigma$, the height of the peak $W_{max}$ and the shape of the fitness peak $Q$ were inferred with more accuracy. For all parameters, the regression and neural network algorithms improved the accuracy of inference relative to the rejection algorithm, and the neural network algorithm was most often the best (Table 2, compare the "rej", "reg" and "nn" columns for each parameter).

With the summary statistics we chose, the number of mutations that were combined together did not affect much the quality of inference. The experimental design with 32 genotypes made of all combinations of 5 mutations performed similarly to the one made of 8 mutations and single and double mutants only (28 genotypes) (Table 2). The design where 20 random mutations were chosen (landscapes B1-B10) did not perform particularly better than the one with 8 mutations and all single and double mutants (29 genotypes in total).

The protocol used to isolate mutations was of critical importance to the quality of inference (Fig. 2, Table 2). Generally, selected mutations allowed the most accurate inference (compare "random", "independently selected" and "co-selected" lines for a given experimental design). In these simulations, the protocol where the four largest mutations were isolated among 48 independently selected mutations performed best and allowed fairly precise inference of the size of mutations (error = 0.145), height of the peak (error = 0.068), and the shape of the peak (error = 0.045) under the neural network algorithm. Protocols that performed best regarding inference of the height and shape of the fitness peak allow a better exploration of the underlying fitness landscape around the fitness optimum. Independently selected mutations and particularly large effect mutations create genotypes that are more likely to be around the fitness peak, especially when genotypes with more than two mutations are included. In contrast, genotypes constructed with random mutations do not always approach the fitness peak and may be confined to relatively linear and uninformative zones of the underlying fitness landscape.

*Parameter inference in experimental datasets*

We obtained the posterior distribution of fitness landscape parameters in the 25 datasets. We used the ABC protocol based on summary statistics and the neural network algorithm, which was shown to work best (Table 2). Note that the neural network algorithm, in rare occasions, resulted in parameters estimates with biologically meaningless values, for example negative values of dimensionality or maximal fitness. This is a known problem (Sünnaker et al. 2013), that happens when none of the summary statistics are very close to the data, such that the neural network regression extrapolates and yields posterior values outside the range of the prior. Results are similar, but the posterior distributions are wider, when using inference based on the full set of fitness values and/or the rejection algorithm.

First, as expected from cross-validation, the posterior distributions were broader for parameters describing dimensionality and shape of the peak (Fig. 3, Table 3). Each empirical landscape could have been generated under a diversity of underlying fitness landscapes. In spite of the uncertainty in parameters, different biological systems exhibited different type of fitness landscapes (Fig. 3).

Three of the experimental systems that were represented by several non-independent empirical landscapes resulted in similar posterior distributions across these "replicated" landscapes. This demonstrates the robustness of the ABC method to slight variation in the set of mutations, to variation in the fitness measure, and to experimental error. For *Aspergillus niger* (Fig. 3, first row), two empirical landscapes A1 and A2 were constructed using two partially overlapping sets of mutations (de Visser et al. 1997). For *Drosophila melanogaster* (Fig. 3, first row), the two landscapes C1 and C2 corresponded to two correlated fitness measures, "productivity" (a measure of lifetime reproductive success) and "mating success" (Whitlock & Bourguet 2000). The posterior distributions of these two landscapes were overlapping, had the same covariance structure, and the median posterior distributions were similar. H1 and H2, two cefotaxime resistance landscapes composed of the same mutations but with replicate MIC measurements, also had similar posterior distribution of parameters (Table 3).

Remarkably, independent empirical landscapes representing the same biological system had similar posterior distribution of parameters. The 10 independent empirical landscapes extracted from the large yeast gene deletion dataset B1-B10 (Costanzo et al. 2010) gave similar posterior distributions characterized in particular by mutations of small effect and a low maximal fitness. The two empirical landscapes of vesicular stomatitis virus, E1 and E2, had extremely similar posterior distribution of parameters, although they had very different statistical properties (Table S1). Different statistical properties arise because of differences in protocol: E1 is composed of independently selected mutations while E2 is composed of random mutations. The fact that we recover similar underlying landscapes for E1 and E2 illustrates the ability of our method to correct for variation due to protocol.

Lastly, underlying landscapes were similar when using independent empirical landscapes obtained in similar biological systems, as revealed by the comparison of the two empirical landscapes of virus on their host (D, E) and of the two landscapes of bacteria adapting to a novel environment (F, G; Fig. 3, third row). In each biological system the two landscapes represented independent experiments; yet posteriors were similar in their marginal distributions and bivariate correlation structure, revealing similar underlying fitness landscapes. The landscape of resistance to pyrimethamine was also quite distinct, with large effect mutations, large maximal fitness and a flat peak (I3; Fig. 3, fourth row).

*Posterior predictive checks: are experimental landscapes compatible with Fisher's model?*

We tested whether the empirical landscapes we analysed were compatible with the hypothesis that Fisher's landscape was the underlying model for the empirical data. An informal test consisted in re-simulating using the posterior distribution of parameters and examining how close these re-simulated landscapes were to the data. We verified that re-simulated landscapes are indeed close to the pseudo-data in the cross-validation, i.e. when the true model was Fisher's model (Fig. 4, left panels). For real data, in contrast, the re-simulated fitness were close to the true fitness for some, but not all, landscapes (Fig. 4, middle panel). More formally, we computed a p-value that expresses the probability that the distance between observed data and simulated datasets would occur if data followed Fisher's model, as described in the methods section (Fig. 4, right panel). We computed this p-value both for the distance based on the full set of fitness values, and for the distance based on summary statistics. The test of rejection based on summary statistics tests whether Fisher's model can reproduce several average statistical properties of landscapes (mean and variance of selection and epistasis coefficient, etc.). The test of rejection based on the full set of fitness values tests whether Fisher's model can reproduce the whole of the data, including specific relationships between genotypes and fitness values not captured by summary statistics. Thus, the test based on the full set of fitness values will be a stronger test of the adequacy of Fisher's model and will reject Fisher's model more often than the test based on summary statistics, because it conserves all information in the landscape.

Fisher's model reproduced the overall statistical properties of all empirical landscapes, but in 6 cases out of 9 it could not reproduce the full structure of empirical landscapes (Table 3). The p-values based on summary statistics were almost always greater than 0.05 (Table 3) (except for MIC landscapes H1-H4 and I1-I2, as discussed in Supplementary Information, Fig. S2). This indicates that the statistical properties of fitness landscapes described by the six summary statistics – mean and variance of epistasis and selection, correlation between epistasis and background fitness, maximal fitness – could be reproduced by Fisher's model. However, Fisher's model was not able to explain fully the structure of 6 fitness landscapes out of 9 (the landscapes B, C, E, F, I3, with p-values smaller than 0.05 in Table 3). We did not identify a single reason why Fisher's model was rejected, but it was often related to mutations with strong negative or positive epistasis (Fig. 5). Fisher's model could reproduce fully only the landscapes of *Aspergillus niger* (A1-A2), of a bacteriophage adapting to its host (D), and of bacteria adapting to a methanol environment (G) (Table 3). In one of the landscapes compatible with Fisher's model, the four beneficial mutations interacted almost additively (Fig. 5, G); but a very different landscape, that includes beneficial and deleterious mutations, and substantial sign epistasis among these, was also compatible with Fisher's model (Fig. 5, A1). In contrast, landscape C1, which looks superficially similar to A1, rejected Fisher's model. Landscape F also rejected Fisher's model, one reason being that the third mutation had very strong positive epistasis with the first mutation. The landscape of pyrimethamine resistance I3 rejected Fisher' model because of two cases of strong reciprocal sign epistasis. Thus, although Fisher's model appears valuable to predict statistical properties of landscapes, in a number of cases it could not explain more detailed properties of experimental landscapes such as mutations presenting large positive or negative epistasis.

In summary, our framework revealed biological differences between the underlying fitness landscapes of 26 experimental landscapes representing 9 independent systems. Fisher's model was generally able to reproduce the statistical properties of empirical landscapes, but not their full structure. In particular only 3 biological systems out of 9 (A, D, G), featuring both very smooth and additive landscapes and more rugged ones, had a structure that was reproduced by Fisher's model.

**Discussion**

Our understanding of the structure of fitness landscapes has greatly improved, in particular thanks to experiments that identify mutations and systematically measure the fitness of a set of genotypes bearing combinations of these mutations. Yet the generality of insights drawn from these empirical landscapes has recently been questioned (Szendro et al. 2013; Schenk et al. 2013; Blanquart et al. 2014). Our analysis shows that the properties of empirical landscapes are heavily dependent on the particular mutations that are sampled (a small number, among a myriad of available mutations), and on the protocol used to identify mutations. We developed a novel framework, based on Approximate Bayesian Computation, to address these challenges and unravel the properties of the underlying fitness landscapes. More precisely, we inferred the underlying fitness landscape, parameterized with Fisher's model, while accounting for the effects of the protocol on the empirical landscapes, and quantifying the uncertainty due to sampling of a limited number of mutations. We used this statistical approach to conduct a survey of fitness landscapes across various species and ecological contexts.

*Summary of the results*

Empirical landscapes, because they are composed of a small number of mutations, generally conveyed limited information on the underlying fitness landscape. This lack of information is manifest in wide posterior distributions and a low accuracy of inference. In other words quite different underlying fitness landscapes may generate similar empirical landscapes. This relates to a previous study where we showed, conversely, that the same underlying landscape results in a variety of empirical landscapes when multiple sets of mutations are sampled (Blanquart et al. 2014). The fact that empirical landscapes are built with a small, and often biased sample of mutations from the underlying fitness landscape suggests that any extrapolation on the global properties of the fitness landscape from measurement on small empirical landscapes should be taken with extreme caution.

While the size of mutations, the height of the peak (maximal fitness) and the shape of the peak were well inferred under some protocols, the number of dimensions under selection was not inferred with accuracy. Importantly, mutations independently selected in several replicates conveyed most information on the underlying fitness landscape, because these allowed an exploration of the most informative regions of the underlying landscapes. With a protocol that included as little as four mutations and all 16 possible genotypes carrying these four mutations, the size of mutations, the height and shape of the peak were well inferred (Table 2).

Fisher's model did not accurately reproduce empirical landscapes in 6 biological systems out of the 9 tested. The conceptual simplicity of Fisher's model and its capacity, so far, to reproduce several experimental observations have made it a popular model to interpret experimental data and generate theoretical predictions (Tenaillon 2014). Fisher's model has been successfully used before to predict the distribution of epistasis coefficients (Martin et al. 2007). Fisher's model also generates sign epistasis, by optimum overshooting, when the ancestral strain is close to the optimum, or by pleiotropic effects, when two mutations have small positive fitness effects (Blanquart et al. 2014). We suggest here that although Fisher's model is able to reproduce several statistical properties of fitness landscapes, it cannot account for their full structure in many cases. This leads to rejection of Fisher's model even with datasets of modest size. Fisher's model could not explain (i) sign epistasis far from the optimum (A1, I3, Fig. 5), (ii) large negative or positive epistasis (C1, F, Fig. 5), (iii) the large variance in selection coefficients and double mutants fitness (B, E). It will be interesting to see whether these patterns can be explained by alternative phenotypic models that allow for some asymmetry around fitness peaks, restricted pleiotropy (mutations affect only a subset of the phenotypes) or anisotropy (mutations do not affect all trait to the same extent).

*Relationship with previous studies*

To our knowledge, only three studies so far have attempted to compare properties of empirical landscapes across species. Szendro et al. (Szendro et al. 2013) quantified ruggedness for 10 experimental landscapes using a set of summary statistics. They showed that experimental levels of ruggedness are similar to those obtained with simulations of simple landscapes made of an additive component and random noise ("Rough Mount Fuji" landscapes). They noticed the strong effect of the experimental protocol on the experimental landscape, and in particular that co-selected mutations tend to produce smoother empirical landscapes. However their framework did not allow disentangling sampling variation due to protocol from variation due to genuine biological differences between systems. Weinreich et al. (2013) analysed 14 empirical landscapes, defined higher-order epistasis coefficients and showed that these coefficients make an important contribution to fitness in all experimental landscapes. Lastly, Weinreich and Knies (2013) fitted

Fisher's model to 7 published datasets using an elegant geometric interpretation of the relationship between the epistasis and selection coefficients. They found Fisher's model fits the data poorly. However it is not clear whether this was due to the data itself, or to the very strong assumptions on which the analytical approach was based: the ancestral strain was always assumed to be perfectly adapted as it was set at the fitness optimum, and all mutations were considered random so that the biasing effects of selection were not accounted for.

Some of the landscapes analysed here have been previously analysed with Fisher's model or similar phenotypic landscapes. Martin et al. (Martin et al. 2007) inferred the parameters of Fisher's model from the distribution of selection coefficient and epistasis coefficient in an RNA virus (our dataset E) (Sanjuán et al. 2004). They found that the distribution of epistasis coefficients is approximately normal with a variance twice that of the variance of the distribution of selection coefficient, in agreement with theoretical predictions from Fisher's model, when the ancestral strain is close to the optimum (Blanquart et al 2014). Accordingly we found that statistical properties of this landscape could be reproduced by Fisher's model, but not its full structure. Last, the yeast deletion dataset (B1-B10) also rejected Fisher's model, as previously reported using a different analysis (Velenich & Gore 2013).

Several studies have attempted to fit phenotypic landscapes to data (Chou et al. 2011; Rokyta et al. 2011; Schenk et al. 2013). In those studies, the underlying phenotypic effects of mutations are considered as parameters that are explicitly estimated and the mapping of phenotypes to fitness is defined by a function (e.g., a Gaussian, or a gamma function). This makes it easier to derive the likelihood, but prevents the use of multivariate landscapes that require a number of parameters proportional to the number of dimensions. Explicitly estimating phenotypes of individual mutations gives interesting insights in the system when the underlying phenotypes are biologically meaningful, and sometimes even measurable. It is also useful if one wants to predict the fitness of combinations of mutations not present in the data. However it requires many parameters even for a simple univariate phenotypic landscape: for example, in dataset D (Rokyta et al. 2011), a univariate-gamma landscape includes 14 parameters while Fisher's model has only 2, and both models perform similarly in terms of AIC. Fisher's model is a useful heuristics to make predictions on the statistical properties of fitness landscapes, but the precise value of the underlying phenotypes is less interesting in such an abstract model.

*Current challenges in the analysis of genotypic fitness landscapes*

In this study we address a number of challenges to fit Fisher's model to a diversity of experimental landscapes. But several other challenges remain to improve our understanding of fitness landscape across species and environments.

(i) Modelling the effects of the protocol on the experimental fitness landscape to infer properly the underling fitness landscape. Here selection was modelled using the "Strong Selection, Weak Mutation" approximation, which is valid when adaptation proceeds by successive invasions of rare beneficial mutants. This approximation was necessary to enable fast simulations required by the ABC approach. However, in some situations of interest in experimental evolution, multiple beneficial mutations compete simultaneously in the population (clonal interference); under this regime beneficial mutations of larger effect tends to invade the population (Nagel et al. 2012). Clonal interference may be important in particular for the landscapes where mutations evolved in the context of experimental evolution (D, E, F, G). The fitness values reported also need to be ecologically relevant, in the sense that they can be used to predict the fate of new mutations competing with the ancestral strain. Exponential growth rates, as reported in many studies, fulfil this condition. But other fitness measures are more dubious. For example in drug resistance landscapes, the fitness measure is commonly the Minimum Inhibitory Concentration. We showed in the example of pyrimethamine resistance that the fitness landscape was quite different when a more correct fitness measure, the growth rate at a given drug concentration, was used. This invites to caution when analysing MIC landscapes from an evolutionary perspective.

(ii) Fitting larger empirical landscapes: Empirical landscapes contain little information on the parameters of their respective underlying landscape. Larger datasets (Costanzo et al. 2010; Hietpas et al. 2011; Bank et al. 2015) may allow more accurate inference and will become much more common in the future. Our ABC method is too computationally intensive to handle such large datasets. New theoretical developments and new statistical techniques need to be developed. These must take into account the potential biases inherent to the data production procedure. A likelihood approach would be ideal, but unfortunately the probability of observing a set of fitness values under Fisher's model is hard to compute as soon as genotypes carry two mutations or more, let alone when mutations were obtained using complex protocols. In essence this is because computing the probability of a fitness value requires integration over all possible values of the unobserved phenotypes.

(iii) Fitting other type of data: Other type of data may prove more informative than empirical landscapes. For example Martin and Lenormand (Martin & Lenormand 2006) use the fitness effects of mutations across environments to infer very precisely the shape of the fitness peak (that is our *Q* parameter), which they find to be very close to *Q=2* (the Gaussian function). Perfeito et al. (Perfeito et al. 2014) show that temporal dynamics of fitness in experimental population allows good inference of the underlying fitness landscape, including dimensionality which is very hard to infer from genotypic landscapes. Again, new theoretical developments may reveal what type of empirical data informs best on the underlying fitness landscape.

*Conclusion*

We have developed a rigorous statistical framework based on Fisher's model to infer the properties of the underlying fitness landscape from empirical landscapes. This framework differs conceptually from previous approaches, as it considers an empirical landscape as a small sample in the vast space of all possible genotypes. This new approach reveals that most experimental protocols reconstruct small landscapes that carry limited information on the true underlying landscape. As a consequence, any analysis and interpretation of empirical landscapes must be embedded within a proper statistical framework that quantifies the uncertainty on the true landscape. Surprisingly, we find that a very broad class of phenotypic models, that has been successful so far in interpreting experimental data, is unable to explain the structure of most empirical fitness landscapes. Yet phenotypic models represent an interesting venue for future research, as they can represent landscapes of large dimensionality with a small number of parameters, and they are more biologically grounded that direct genotype-fitness maps. Much larger empirical landscapes will become more frequent in the future; a model-based and statistically grounded analysis of these large landscapes will improve our understanding of the structure of fitness landscapes across species and environments.

**Acknowledgements**

We thank Guillaume Achaz, Luis-Miguel Chevin, Florence Débarre, Luca Ferretti, Thomas Lenormand, and Olivier Tenaillon for discussions and useful comments. Comments from Craig Miller, Daniel Weinreich and one anonymous reviewer greatly improved the manuscript. This work was supported by Danish Research Council (FFF-FNU), the European Research Council under the European Union's Seventh Framework Program (ERC Grant 311341 to T.B.) and the Bettencourt Foundation (Young Researcher Award to F.B.).

## Q=0.5, random mutations

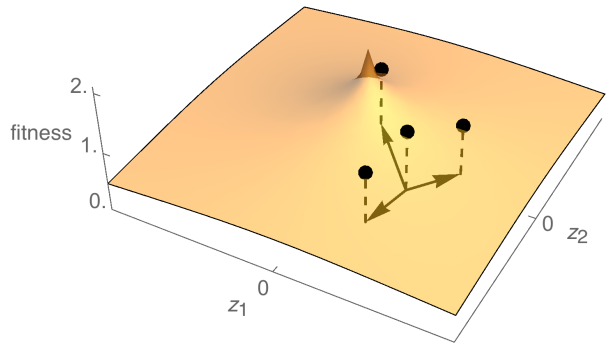 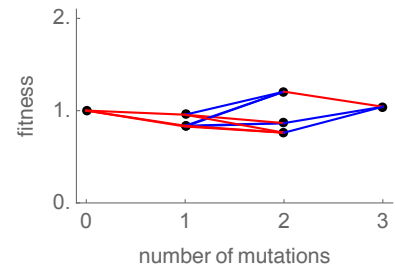

## Q=2, co-selected mutations

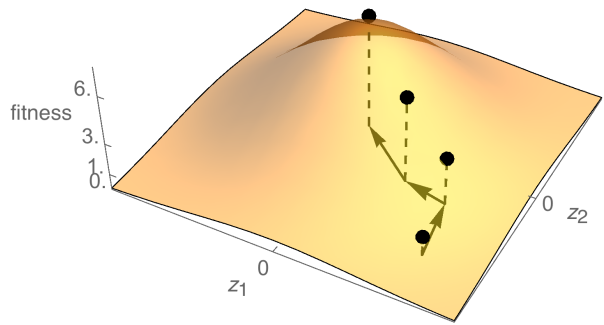 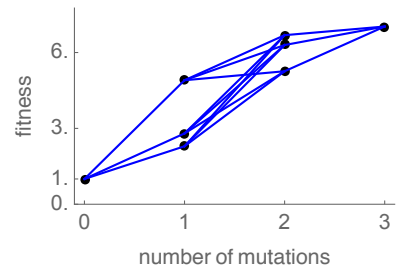

## Q=4, independently selected mutations

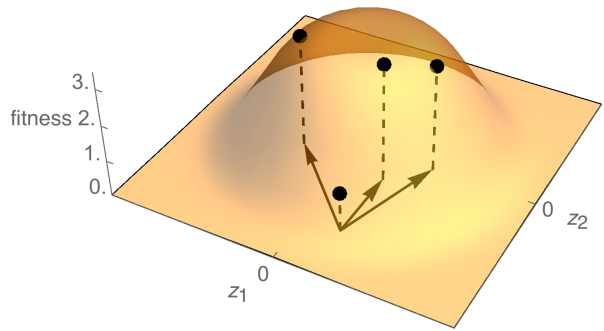 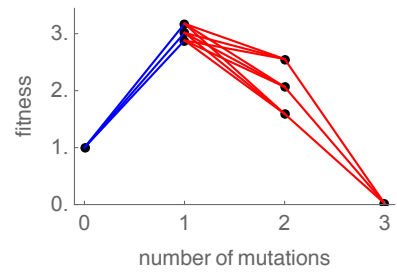

Figure 1: A diversity of genotypic landscapes can be generated by Fisher's fitness landscape model. Each row shows an example of Fisher's landscape, with three mutations depicted as arrows in the phenotypic space (left), and the empirical landscape resulting from these mutations in combination (i.e., 8 genotypes) (right). Blue edges denote mutations that are beneficial in the considered background, while red edges denote deleterious mutations. Top row: a sharp landscape with $Q=0.5$, and where the three mutations are random mutations. Middle row: Fisher's classic landscape with $Q=2$, and three co-selected mutations. Bottom row: $Q=4$, and three independently selected mutations.

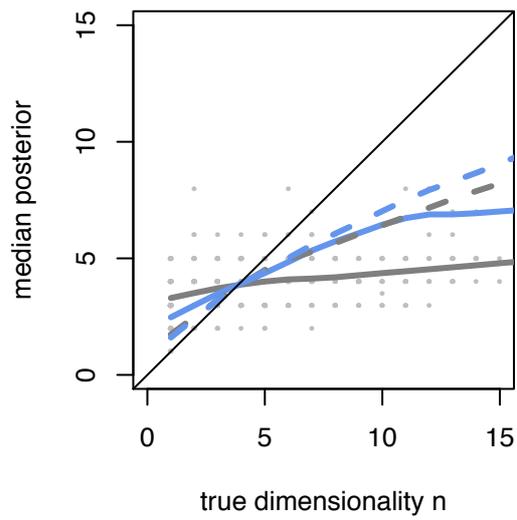
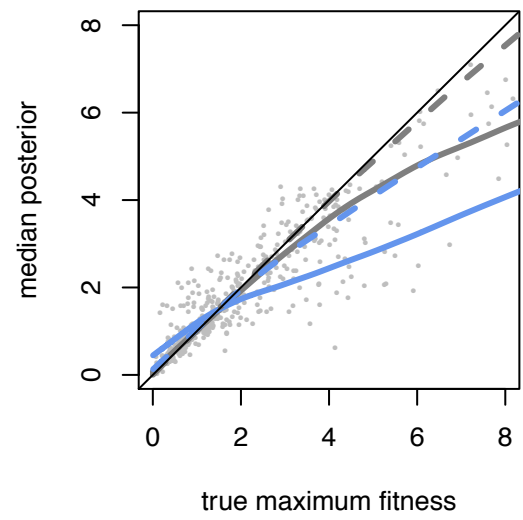
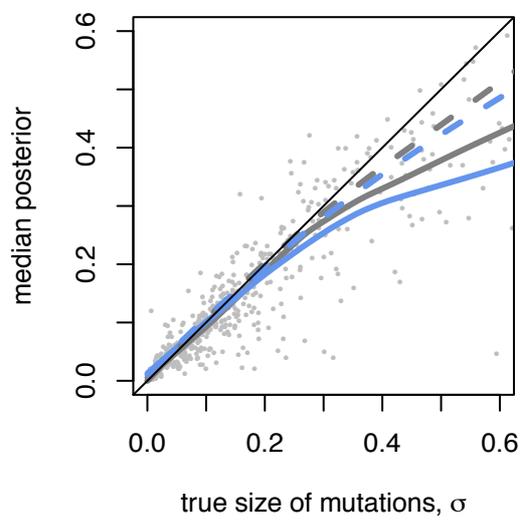
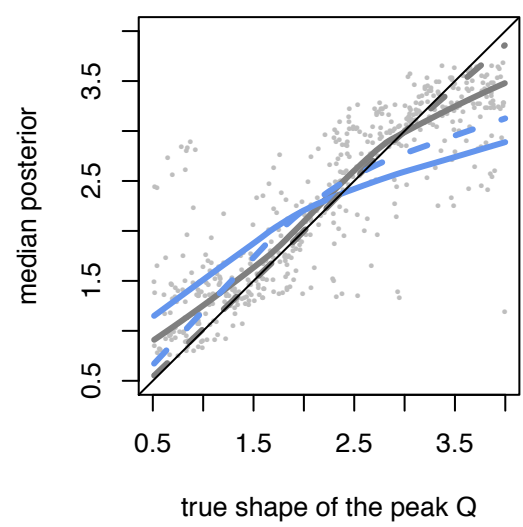

Figure 2: Accuracy of inference for different methods and different datasets. The median posterior distribution for the rejection algorithm is shown as a function of the true parameter for each of the 500 cross-validation datasets (grey points), when the set of genotypes is composed of all combinations of 4 independently selected mutations, chosen as the four largest effect mutations among a set of 48 mutations, as in H4 (Schenk et al. 2013). Perfect inference corresponds to all points on the y = x line. For clarity, we represent this cloud of points with a local non-linear fit (grey line). The equivalent linear fit for the neural network algorithm is shown as a grey dashed line. The plain and dashed blue line similarly show the local linear fit for rejection and neural network algorithms, for the dataset composed of 20 random mutations, and single and double mutants only (as in B1-B10). The neural network algorithm generally improves inference compared to the rejection algorithm. The dataset composed of all combinations of 4 selected mutations performs better than the one composed of 20 random mutations and single and double mutants.

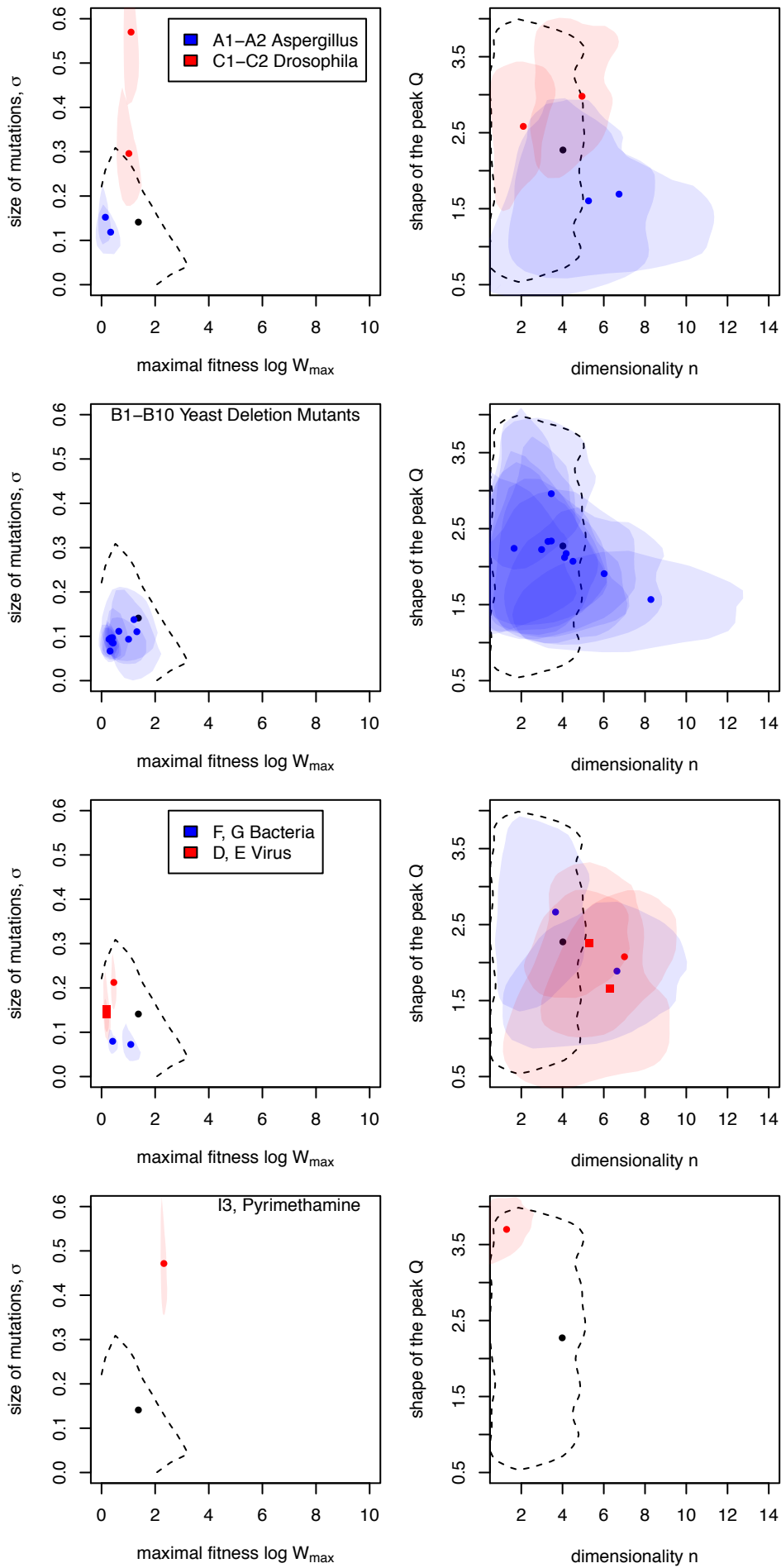

Figure 3: Posterior distribution of parameters for all experimental landscapes. From top to bottom: A1-A2 (Aspergillus) and C1-C2 (Drosophila); the yeast deletion dataset (B1-B10); virus evolving on their host (D, E1, E2) and bacteria in a novel medium (F, G); adaptation to an environment containing pyrimethamine (I3). The black point shows the median of the prior and the dashed line delineates the 50% higher density region. The points show the median of the posteriors and the shaded areas show the 50% higher posterior density regions for the datasets.

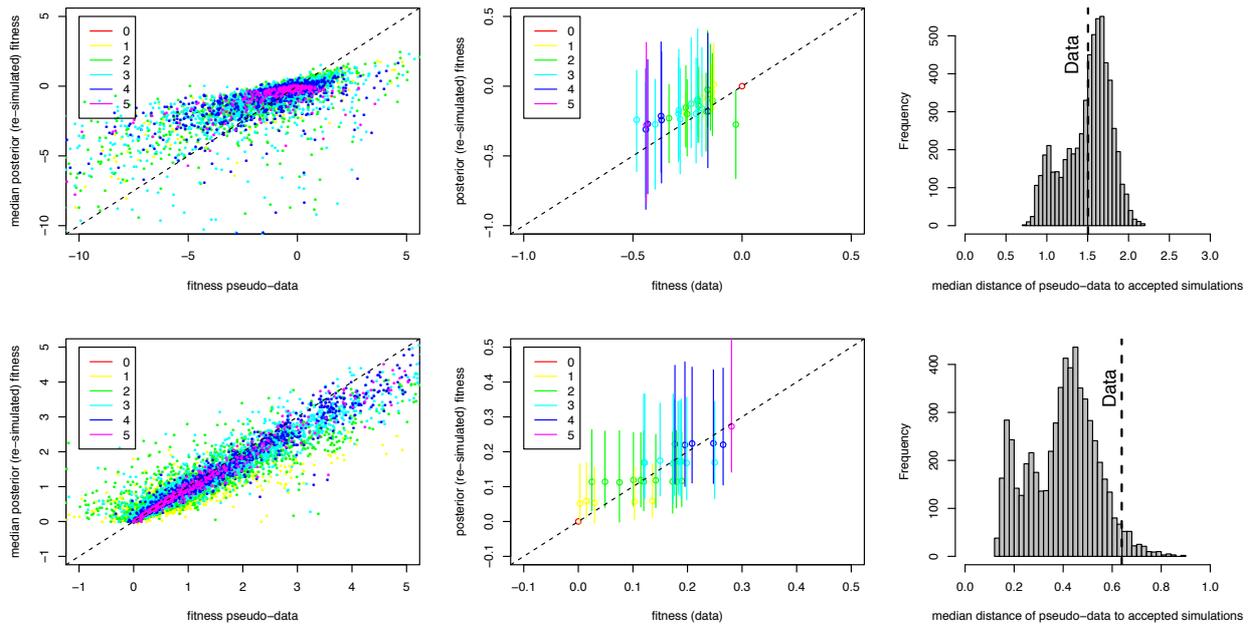

Figure 4: Posterior predictive checks on two example datasets. One dataset is compatible with Fisher's model (top row; *Aspergillus* dataset A1) and one rejects Fisher's model (bottom row, F). Left panel: the median posterior fitness against the "true" fitness of pseudo-data generated under Fisher's model for the cross-validation, showing that when the pseudo-data has been generated using Fisher as the true model, the posterior fitness are close to the true fitness values. Middle panel: posterior predicted log-fitness as a function of the true experimental log-fitness. The point is the median posterior and the intervals show the 2.5%-97.5% interval. The colour code indicates the number of mutations of each genotype, the ancestor in red being set to log-fitness = 0. The median posterior fitnesses are very well correlated with the true fitnesses when the landscape is compatible with Fisher's model, but less so when Fisher's model is rejected. Right panel: the median distance of pseudo-data to the accepted simulations, when the pseudo-data is simulated under Fisher's model using the posterior parameters. This distribution together with the observed median distance for the experimental data (dashed line) is used to calculate the p-value corresponding to the null hypothesis "the underlying fitness landscape is Fisher's model".

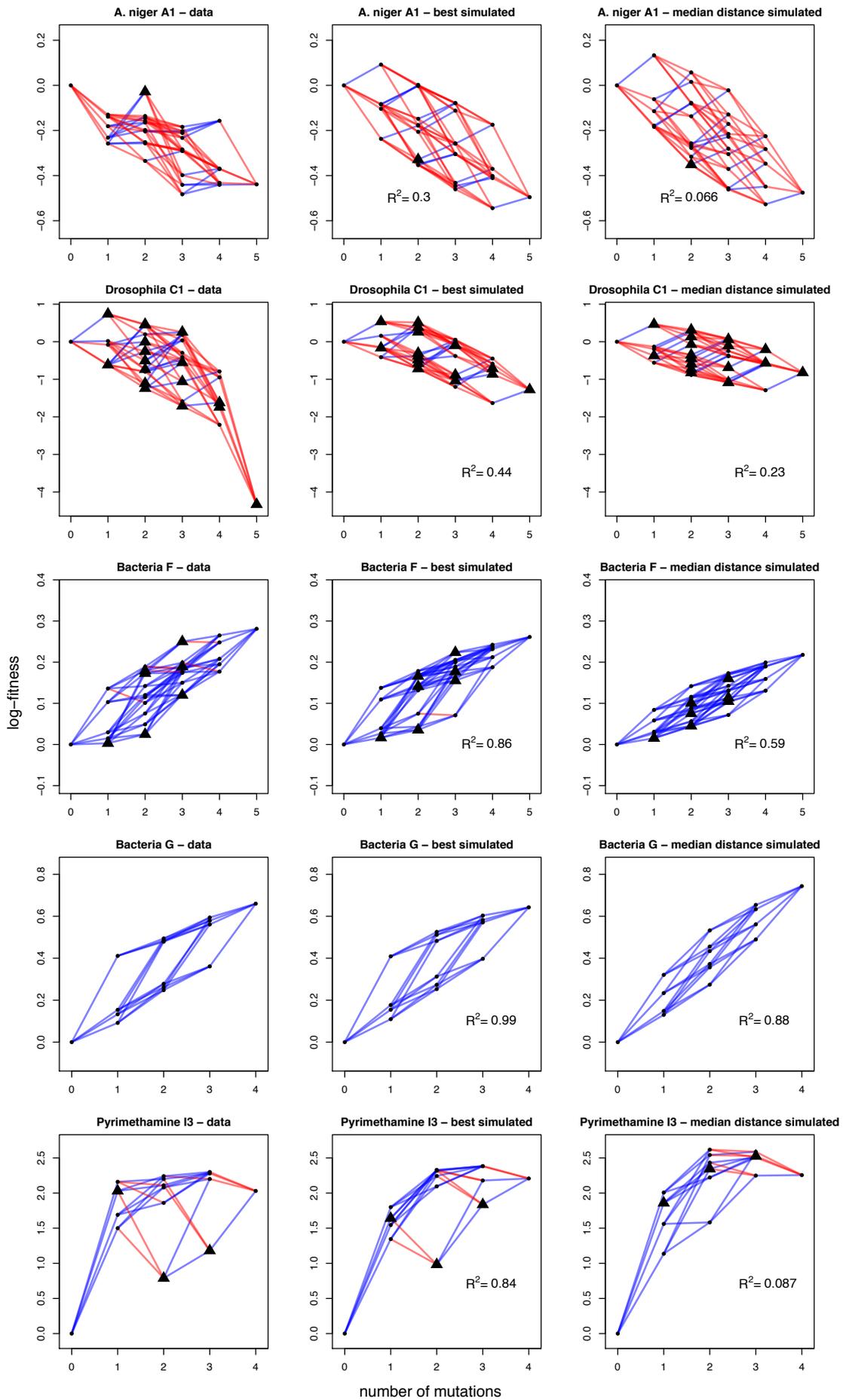

Figure 5: Empirical landscapes compared with simulated landscapes. For each dataset, the data (left) is shown side by side with the simulated genotypic landscape closest to the data in terms of Euclidean distance (middle), and a typical simulated landscape, defined as the landscape, among all simulated landscapes retained by the ABC framework, whose distance to the data was closest to the median distance. The coefficient of determination $R^2$ is also shown. Blue edges are beneficial mutations, red edges are deleterious mutations. Fitness values that are particularly unexpected under Fisher's model are marked with a triangle.

## Table 1: Summary of datasets

| Name | Species | Environment | Mutation type | Mutation and genotype number | Fitness measure | Measurement error | Note | References |
|---|---|---|---|---|---|---|---|---|
| A1, A2 | *Aspergillus niger* | Minimal medium | Random (phenotypic markers) | 2 datasets of 5 mutations, $2^5$ genotypes | Rate of increase in colony radius per unit time ("radial growth rate"), relative to the ancestor. | We calculated a single standard error using the two replicate measurements. | The two sets of 5 mutations are not independent. | (de Visser et al. 1997; de Visser et al. 2009) |
| B1-B10 | *Saccharomyces cerevisiae* | Standard medium (on plates) | Random mutations (gene deletions) | 1711+3885 mutations, 5.4 million genotypes | Increase in colony size per unit time relative to the ancestor. | Standard error was reported for each fitness measure. | Analysis done on 10 independent subsets of 20 mutations. 1 + 20 + 100 genotypes corresponding to single mutants and double mutants. | (Costanzo et al. 2010) |
| C1, C2 | *Drosophila melanogaster* | Lab environment. | Random (phenotypic markers) | 5 mutations, $2^5$ genotypes | Productivity (product of fecundity and survival) **C1**, and mating success **C2** | We roughly estimated standard error for the productivity measure (SI). | 2 genotypes with 0 mating success removed. | (Whitlock & Bourguet 2000) |
| D | ssDNA bacteriophage ID11 | E. coli (host) | Independently selected (experimental evolution) | 9 mutations, 9 single mutants, 18 double mutants | $\log_2$ increase in phage population per hour | Standard error was reported for each fitness measure. | - | (Rokyta et al. 2011) |
| E1, E2 | *Vesicular stomatitis virus* | Baby hamster kidney (BHK21) cells (host) | Independently selected (**E1**, found in natural isolates) and random (**E2**) | 6 mutations, 6 single mutants, 15 double mutants (**E1**) 28 mutations, 76 double mutants (**E2**) | Growth rate relative to ancestor | Standard error was reported for each fitness measure. | - | (Sanjuán et al. 2004) |
| F | *Escherichia coli* | New, Low-glucose environment | Co-selected in experimental evolution | 5 mutations, $2^5$ genotypes | Growth rate relative to ancestor | 95% CI were reported. | - | (Khan et al. 2011) |
| G | *Methylobacterium extorquens* (genetically modified strain) | Methanol environment | Co-selected in experimental evolution | 4 mutations, $2^4$ genotypes | Growth rate relative to ancestor | Standard errors were not reported, but we estimated them using standard errors reported for another set of genotypes (SI). | Order of fixation of mutations not known. Mutations were assumed to fix from the largest effect mutation to the smallest effect mutation. | (Chou et al. 2011) |
| H1, H2 | *Escherichia coli* | Cefotaxime (β lactam antibiotic) | Independently selected (found in natural isolates). Mutations chosen because together increase resistance to cefotaxime 100, 000 fold. | 5 mutations, $2^5$ genotypes | Cefotaxime resistance measured as Minimum Inhibitory Concentration | For **H1**, we calculated single errors using the three replicate measurements. For **H2**, standard errors were reported. | **H2** is the same dataset as **H1**, with MIC re-measured on same genotypes. Resistance to piperacillin+clavulanic acid was also measured but not used here. | (Weinreich et al. 2006) (Tan et al. 2011) |
| H3, H4 | *Escherichia coli* | Cefotaxime (β lactam antibiotic) | Independently selected (found in natural isolates) | 4 mutations, $2^4$ genotypes. Two independent datasets | Cefotaxime resistance measured as $IC_{99.99}$ (highly correlated with MIC). | Standard errors were not reported, but we used the average standard error of **H2**. | One dataset with 4 mutations of smallest effect **H3**, one with 4 mutations of largest effect **H4** among 48 mutations. | (Schenk et al. 2013) |
| I1, I2, I3 | *Plasmodium falciparum* DHFR gene transformed into E. coli **I1** and S. cerevisiae **I2**. *Plasmodium vivax* DHFR gene transformed into S. cerevisiae (**I3**) | Pyrimethamine (antimalarial drug) | Independently selected (found in clinical isolates) | 4 mutations, $2^4$ genotypes **I2** includes the same 4 mutations as **I1** plus 2 additional mutations affecting another locus. | Pyrimethamine resistance measured as $IC_{50}$ in μg/mL (**I1**) and M (mol/L) (**I2**). Growth rate of the transformed strain at concentration 1 μmol/L (**I3**). | Standard errors for **I1** and **I2** were reported. | In **I2** (Brown et al. 2010), one locus has three possible alleles. The third allele was ignored, resulting in 17 fitness values. | (Lozovsky et al. 2009) (Brown et al. 2010) (Jiang et al. 2013) |

**Table 2: Expected prediction error under various experimental designs**

| Experimental design | Type of mutation | Landscapes using this protocol | n | | | W_max | | | σ | | | Q | | |
|---|---|---|---|---|---|---|---|---|---|---|---|---|---|---|
| | | | rej | reg | nn | rej | reg | nn | rej | reg | nn | rej | reg | nn |
| 5 mutations, $2^5$ genotypes | random | A, C | 0.849 | 0.685 | 0.641 | 0.572 | 0.391 | 0.37 | 0.439 | 0.347 | 0.32 | 0.672 | 0.491 | 0.429 |
| | independently selected | - | 0.912 | 0.787 | 0.702 | 0.345 | 0.201 | 0.18 | 0.332 | 0.192 | **0.185** | 0.35 | 0.147 | **0.092** |
| | co-selected | F | 0.831 | 0.733 | 0.625 | 0.343 | 0.193 | 0.165 | 0.53 | 0.367 | 0.332 | 0.392 | 0.236 | 0.171 |
| 4 mutations, $2^4$ genotypes | random | - | 0.925 | 0.803 | 0.782 | 0.794 | 0.575 | 0.544 | 0.425 | 0.358 | 0.351 | 0.642 | 0.524 | 0.447 |
| | independently selected | I | 0.867 | 0.763 | 0.668 | 0.41 | 0.216 | 0.184 | 0.43 | 0.254 | 0.23 | 0.482 | 0.266 | 0.2 |
| | co-selected | G | 0.9 | 0.76 | 0.694 | 0.373 | 0.197 | 0.165 | 0.493 | 0.334 | 0.298 | 0.425 | 0.326 | 0.236 |
| 8 mutations, 8 single and 20 double mutants | random | - | 0.8 | 0.582 | **0.539** | 0.684 | 0.476 | 0.496 | 0.367 | 0.298 | 0.294 | 0.551 | 0.444 | 0.401 |
| | independently selected | - | 0.747 | 0.687 | 0.629 | 0.441 | 0.291 | 0.216 | 0.406 | 0.294 | 0.281 | 0.466 | 0.33 | 0.29 |
| | co-selected | - | 0.765 | 0.671 | 0.617 | 0.402 | 0.215 | 0.175 | 0.427 | 0.263 | 0.231 | 0.267 | 0.229 | 0.2 |
| 20 mutations, up to 121 genotypes | random | B | 0.725 | 0.625 | **0.536** | 0.475 | 0.253 | 0.207 | 0.355 | 0.232 | 0.221 | 0.618 | 0.449 | 0.389 |
| 9 mutations, 9 single mutants, 18 double mutants | independently selected | D | 0.744 | 0.676 | 0.634 | 0.366 | 0.221 | **0.147** | 0.379 | 0.268 | 0.243 | 0.452 | 0.366 | 0.318 |
| 6 mutations, 6 single mutants, 15 double mutants | independently selected | E | 0.812 | 0.799 | 0.761 | 0.448 | 0.245 | 0.182 | 0.387 | 0.33 | 0.301 | 0.668 | 0.588 | 0.545 |
| 5 mutations, $2^5$ genotypes | independently selected, high fitness combination | H1, H2 | 0.855 | 0.745 | 0.608 | 0.238 | 0.086 | **0.063** | 0.415 | 0.246 | **0.215** | 0.236 | 0.097 | **0.049** |
| 4 mutations, $2^4$ genotypes | independently selected, small fitness effect mutants | H3 | 0.801 | 0.838 | 0.778 | 0.687 | 0.52 | 0.433 | 0.499 | 0.406 | 0.379 | 0.687 | 0.477 | 0.361 |
| 4 mutations, $2^4$ genotypes | independently selected, large fitness effect mutants | H4 | 0.938 | 0.72 | **0.556** | 0.258 | 0.1 | **0.068** | 0.249 | 0.149 | **0.145** | 0.274 | 0.099 | **0.045** |

Prediction error for the four parameters of Fisher's model, for several experimental designs (based on single and double mutants, or complete sets of mutations and all associated genotypes) and selection procedures (random, independently selected, co-selected mutations), when the 6 summary statistics were used in the ABC algorithm. For each parameter, the three lowest prediction errors are in bold, highlighting the protocol and inference algorithms that perform best.

**Table 3: Posterior distribution of parameters and posterior predictive checks, neural network algorithm**

| Reference | Name | Neural network algorithm | | | | p-value (summary) | p-value (full) |
|---|---|---|---|---|---|---|---|
| | | n | $W_{max}$ | $\sigma$ | Q | | |
| - | prior | 4 (1; 19) | 1.39 (0.05; 7.39) | 0.14 (0.01; 0.74) | 2.25 (0.59; 3.91) | | |
| (de Visser et al. 1997) | A1 | 5.24 (0.59; 20.96) | 0.14 (-0.07; 1.88) | 0.15 (0.08; 0.37) | 1.60 (0.42; 3.52) | **0.17** | **0.83** |
| | A2 | 6.72 (1.63; 23.09) | 0.34 (-0.09; 3.12) | 0.12 (0.05; 0.31) | 1.69 (0.66; 3.51) | **0.23** | **0.97** |
| (Costanzo et al. 2010) | B1 | 6.00 (1.54; 19.08) | 1.01 (0.26; 3.88) | 0.09 (0.04; 0.29) | 1.91 (0.91; 3.75) | **0.25** | 0 |
| | B2 | 3.44 (0.31; 12.82) | 0.28 (0.12; 1.02) | 0.09 (0.05; 0.20) | 2.96 (1.64; 4.19) | **0.3** | 0.02 |
| | B3 | 3.33 (0.13; 13.78) | 0.40 (0.13; 1.78) | 0.10 (0.06; 0.23) | 2.33 (1.19; 4.36) | **0.31** | 0.01 |
| | B4 | 8.28 (1.64; 24.06) | 1.21 (0.04; 5.10) | 0.14 (0.03; 0.48) | 1.57 (0.77; 3.13) | **0.16** | **0.06** |
| | B5 | 4.16 (0.89; 14.84) | 0.43 (0.07; 2.37) | 0.08 (0.05; 0.23) | 2.17 (1.02; 4.37) | **0.26** | 0.02 |
| | B6 | 3.01 (-0.72; 13.87) | 0.34 (-0.05; 1.96) | 0.10 (0.05; 0.29) | 2.23 (1.16; 4.68) | **0.24** | 0.01 |
| | B7 | 4.47 (-0.25; 15.71) | 0.64 (0.12; 2.23) | 0.11 (0.05; 0.33) | 2.07 (0.97; 4.12) | **0.14** | 0.02 |
| | B8 | 1.63 (-1.91; 12.29) | 1.32 (0.32; 4.92) | 0.11 (0.00; 0.48) | 2.24 (1.10; 4.17) | 0.02 | 0.01 |
| | B9 | 4.12 (0.48; 15.58) | 0.39 (0.02; 2.38) | 0.09 (0.04; 0.26) | 2.12 (1.07; 4.26) | **0.17** | 0 |
| | B10 | 3.46 (0.63; 15.18) | 0.32 (0.07; 1.58) | 0.07 (0.04; 0.20) | 2.34 (1.08; 4.35) | **0.06** | 0.01 |
| (Whitlock & Bourguet 2000) | C1 | 4.92 (2.12; 13.24) | 1.02 (0.58; 3.20) | 0.30 (0.16; 0.66) | 2.98 (1.71; 4.06) | **0.05** | 0 |
| | C2 | 2.09 (0.21; 7.03) | 1.10 (0.82; 2.38) | 0.57 (0.39; 1.03) | 2.58 (1.01; 3.57) | **0.06** | 0 |
| (Rokyta et al. 2011) | D | 7.00 (2.95; 15.21) | 0.46 (0.36; 0.82) | 0.21 (0.15; 0.39) | 2.08 (0.83; 3.82) | **0.15** | **0.08** |
| (Sanjuán et al. 2004) | E1 | 6.28 (1.64; 19.82) | 0.19 (0.06; 0.86) | 0.15 (0.07; 0.41) | 1.65 (0.23; 3.79) | **0.37** | 0.01 |
| | E2 | 5.28 (2.11; 12.45) | 0.20 (0.09; 0.55) | 0.14 (0.10; 0.25) | 2.26 (1.34; 3.42) | **0.16** | 0.03 |
| (Khan et al. 2011) | F | 6.62 (1.63; 22.28) | 0.42 (0.21; 0.98) | 0.08 (0.05; 0.19) | 1.89 (0.81; 3.70) | **0.43** | 0.03 |
| (Chou et al. 2011) | G | 3.65 (0.86; 15.86) | 1.09 (0.73; 2.48) | 0.07 (0.03; 0.21) | 2.67 (1.30; 4.05) | **0.42** | **0.43** |
| (Weinreich et al. 2006) | H1 | 14.39 (7.25; 29.54) | 12.97 (12.16; 15.73) | 0.89 (0.64; 1.46) | 1.40 (0.14; 2.48) | 0.01 | 0 |
| (Tan et al. 2011) | H2 | 13.18 (5.76; 28.86) | 12.02 (10.87; 14.83) | 0.46 (0.18; 1.08) | 1.83 (0.81; 2.80) | 0.01 | 0 |
| (Schenk et al. 2013) | H3 | 4.81 (1.89; 15.30) | 3.17 (1.08; 8.91) | 0.30 (0.13; 0.79) | 2.94 (1.53; 3.91) | **0.34** | **0.07** |
| | H4 | 8.89 (5.63; 17.44) | 6.24 (5.27; 7.94) | 0.75 (0.51; 1.13) | 1.40 (0.62; 2.15) | 0 | 0 |
| (Lozovsky et al. 2009) | I1 | 8.24 (3.79; 19.68) | 9.20 (7.78; 14.61) | 0.57 (0.26; 1.24) | 2.22 (0.55; 3.51) | 0.02 | 0 |
| (Brown et al. 2010) | I2 | 5.16 (2.50; 13.08) | 7.76 (7.41; 8.95) | 0.23 (0.15; 0.37) | 3.84 (3.17; 4.49) | 0 | 0 |
| (Jiang et al. 2013) | I3 | 1.28 (-0.58; 5.47) | 2.33 (2.19; 2.71) | 0.47 (0.32; 0.79) | 3.70 (3.11; 4.24) | **0.12** | 0.03 |

The median posterior distribution of parameters and the 2.5% - 97.5% quantile interval (equivalent to 95% higher posterior density) of the posterior distribution of parameters for the rejection algorithm. The prior is shown for comparison (first row). The p-value for the test of adequacy with Fisher's model is indicated.

# Epistasis and the structure of fitness landscapes: are experimental fitness landscapes compatible with Fisher's Geometric Model? – Supplementary Information

François Blanquart, Thomas Bataillon

## Details of the datasets

All datasets used in this study were published before, but we reproduce these datasets here. Empirical fitness landscapes are presented as tables where each line is a genotype. The first columns represent the genotypes as a series of 0 and 1 denoting absence or presence of the mutation at each locus. The following columns are, in order, the fitness measure given in the reference, the standard error of this fitness measure, our log-fitness measure, defined as the log of the fitness of each genotype divided by the fitness of the ancestor, and finally the standard error of the log-fitness measure.

## A1

In A1 and A2, fitness was the rate of increase in colony radius per unit time. All radial growth rates were reported relative to that of the ancestor. The absolute radial growth rate of the ancestor was not reported. The data was originally analyzed in (de Visser et al. 1997) and reported in (de Visser et al. 2009), table 1.

| | | | | | | | | |
|---|---|---|---|---|---|---|---|---|
| 0 | 0 | 0 | 0 | 0 | 1     | 0.0687 | 0       | 0.0893 |
| 1 | 0 | 0 | 0 | 0 | 0.878 | 0.0687 | -0.13   | 0.0893 |
| 0 | 1 | 0 | 0 | 0 | 0.834 | 0.0687 | -0.181  | 0.0893 |
| 0 | 0 | 1 | 0 | 0 | 0.87  | 0.0687 | -0.139  | 0.0893 |
| 0 | 0 | 0 | 1 | 0 | 0.772 | 0.0687 | -0.258  | 0.0893 |
| 0 | 0 | 0 | 0 | 1 | 0.793 | 0.0687 | -0.232  | 0.0893 |
| 1 | 1 | 0 | 0 | 0 | 0.865 | 0.0687 | -0.145  | 0.0893 |
| 1 | 0 | 1 | 0 | 0 | 0.854 | 0.0687 | -0.158  | 0.0893 |
| 1 | 0 | 0 | 1 | 0 | 0.773 | 0.0687 | -0.257  | 0.0893 |
| 1 | 0 | 0 | 0 | 1 | 0.873 | 0.0687 | -0.136  | 0.0893 |
| 0 | 1 | 1 | 0 | 0 | 0.816 | 0.0687 | -0.204  | 0.0893 |
| 0 | 1 | 0 | 1 | 0 | 0.716 | 0.0687 | -0.335  | 0.0893 |
| 0 | 1 | 0 | 0 | 1 | 0.848 | 0.0687 | -0.165  | 0.0893 |
| 0 | 0 | 1 | 1 | 0 | 0.778 | 0.0687 | -0.252  | 0.0893 |
| 0 | 0 | 1 | 0 | 1 | 0.82  | 0.0687 | -0.198  | 0.0893 |
| 0 | 0 | 0 | 1 | 1 | 0.972 | 0.0687 | -0.0284 | 0.0893 |
| 1 | 1 | 1 | 0 | 0 | 0.816 | 0.0687 | -0.203  | 0.0893 |
| 1 | 1 | 0 | 1 | 0 | 0.748 | 0.0687 | -0.291  | 0.0893 |
| 1 | 1 | 0 | 0 | 1 | 0.832 | 0.0687 | -0.184  | 0.0893 |
| 1 | 0 | 1 | 1 | 0 | 0.748 | 0.0687 | -0.29   | 0.0893 |
| 1 | 0 | 1 | 0 | 1 | 0.792 | 0.0687 | -0.233  | 0.0893 |
| 1 | 0 | 0 | 1 | 1 | 0.753 | 0.0687 | -0.284  | 0.0893 |
| 0 | 1 | 1 | 1 | 0 | 0.617 | 0.0687 | -0.483  | 0.0893 |
| 0 | 1 | 1 | 0 | 1 | 0.81  | 0.0687 | -0.211  | 0.0893 |
| 0 | 1 | 0 | 1 | 1 | 0.644 | 0.0687 | -0.441  | 0.0893 |
| 0 | 0 | 1 | 1 | 1 | 0.672 | 0.0687 | -0.398  | 0.0893 |
| 1 | 1 | 1 | 1 | 0 | 0.69  | 0.0687 | -0.371  | 0.0893 |
| 1 | 1 | 1 | 0 | 1 | 0.855 | 0.0687 | -0.157  | 0.0893 |
| 1 | 1 | 0 | 1 | 1 | 0.649 | 0.0687 | -0.432  | 0.0893 |
| 1 | 0 | 1 | 1 | 1 | 0.692 | 0.0687 | -0.369  | 0.0893 |
| 0 | 1 | 1 | 1 | 1 | 0.644 | 0.0687 | -0.441  | 0.0893 |
| 1 | 1 | 1 | 1 | 1 | 0.645 | 0.0687 | -0.439  | 0.0893 |

**A2**

| | | | | | | | | |
|---|---|---|---|---|---|---|---|---|
| 0 | 0 | 0 | 0 | 0 | 1 | 0.0687 | 0 | 0.0893 |
| 1 | 0 | 0 | 0 | 0 | 0.878 | 0.0687 | -0.13 | 0.0893 |
| 0 | 1 | 0 | 0 | 0 | 0.834 | 0.0687 | -0.181 | 0.0893 |
| 0 | 0 | 1 | 0 | 0 | 0.87 | 0.0687 | -0.139 | 0.0893 |
| 0 | 0 | 0 | 1 | 0 | 0.908 | 0.0687 | -0.096 | 0.0893 |
| 0 | 0 | 0 | 0 | 1 | 0.772 | 0.0687 | -0.258 | 0.0893 |
| 1 | 1 | 0 | 0 | 0 | 0.865 | 0.0687 | -0.145 | 0.0893 |
| 1 | 0 | 1 | 0 | 0 | 0.854 | 0.0687 | -0.158 | 0.0893 |
| 1 | 0 | 0 | 1 | 0 | 0.924 | 0.0687 | -0.0796 | 0.0893 |
| 1 | 0 | 0 | 0 | 1 | 0.773 | 0.0687 | -0.257 | 0.0893 |
| 0 | 1 | 1 | 0 | 0 | 0.816 | 0.0687 | -0.204 | 0.0893 |
| 0 | 1 | 0 | 1 | 0 | 0.852 | 0.0687 | -0.16 | 0.0893 |
| 0 | 1 | 0 | 0 | 1 | 0.716 | 0.0687 | -0.335 | 0.0893 |
| 0 | 0 | 1 | 1 | 0 | 0.855 | 0.0687 | -0.157 | 0.0893 |
| 0 | 0 | 1 | 0 | 1 | 0.778 | 0.0687 | -0.252 | 0.0893 |
| 0 | 0 | 0 | 1 | 1 | 0.784 | 0.0687 | -0.243 | 0.0893 |
| 1 | 1 | 1 | 0 | 0 | 0.816 | 0.0687 | -0.203 | 0.0893 |
| 1 | 1 | 0 | 1 | 0 | 0.878 | 0.0687 | -0.13 | 0.0893 |
| 1 | 1 | 0 | 0 | 1 | 0.748 | 0.0687 | -0.291 | 0.0893 |
| 1 | 0 | 1 | 1 | 0 | 0.942 | 0.0687 | -0.0603 | 0.0893 |
| 1 | 0 | 1 | 0 | 1 | 0.748 | 0.0687 | -0.29 | 0.0893 |
| 1 | 0 | 0 | 1 | 1 | 0.795 | 0.0687 | -0.229 | 0.0893 |
| 0 | 1 | 1 | 1 | 0 | 0.858 | 0.0687 | -0.153 | 0.0893 |
| 0 | 1 | 1 | 0 | 1 | 0.617 | 0.0687 | -0.483 | 0.0893 |
| 0 | 1 | 0 | 1 | 1 | 0.724 | 0.0687 | -0.323 | 0.0893 |
| 0 | 0 | 1 | 1 | 1 | 0.745 | 0.0687 | -0.294 | 0.0893 |
| 1 | 1 | 1 | 1 | 0 | 0.825 | 0.0687 | -0.192 | 0.0893 |
| 1 | 1 | 1 | 0 | 1 | 0.69 | 0.0687 | -0.371 | 0.0893 |
| 1 | 1 | 0 | 1 | 1 | 0.665 | 0.0687 | -0.408 | 0.0893 |
| 1 | 0 | 1 | 1 | 1 | 0.686 | 0.0687 | -0.376 | 0.0893 |
| 0 | 1 | 1 | 1 | 1 | 0.64 | 0.0687 | -0.446 | 0.0893 |
| 1 | 1 | 1 | 1 | 1 | 0.622 | 0.0687 | -0.474 | 0.0893 |

Each fitness was measured twice, allowing us to estimate the standard error of fitness measurements. There was no indication that measurement error systematically varied with the magnitude of fitness values. Therefore, we used a single measurement error, estimated at $0.097$. This means each fitness value calculated as the average fitness over the two replicates was associated with a measurement error of $\sigma_{error} = 0.097/\sqrt{2} \approx 0.069$. To calculate the standard error on the log scale, we first computed 95% lower and upper bounds on the fitness values as $W \pm 2\sigma_{error}$, which is a good approximation when the error is approximately normally distributed. Then we log-transformed these lower and upper bounds, and computed back the error standard deviation on the log scale as the difference between upper and lower bound divided by 4. This gave an estimated measurement error on the log scale of $\sigma_{error}^{log} = 0.0893$.

**B1-B10**

These landscapes represent 10 random sub-samples of a large dataset (Costanzo et al. 2010). The full dataset is available at http://drygin.ccbr.utoronto.ca/~costanzo2009/

The fitness measure was the increase in colony size per unit time, relative to the ancestral strain. The absolute growth rate of the ancestor was not reported. Each sub-sample included 20 mutations and 121 genotypes (ancestral strain, 20 single mutants, and 100 double mutants). Sub-samples were generated using a custom R code available upon request.

A standard error was reported for each fitness value. We transformed these into standard error on the log scale using the same procedure as described above for A1 and A2 landscapes.

**C1**

The fitness measures were productivity (for C1), which is the number of adult offspring of the strain (thus, it combines fecundity and offspring survival) and mating success (for C2). Both measures were taken in competition with a reference *D. melanogaster* strain carrying a visible mutation, and the resulting number of offspring (or number of matings) was divided by the corresponding number for the control strain. The data is available in (Whitlock & Bourguet 2000), Table 1 (the log fitness is reported).

The error attached to each fitness measure was not reported but we estimated roughly standard error for the productivity assay. For this assay, there were on average 52.5 replicates, each of them including three mated females of the genotype of interest in competition with three mated females of the reference strain. Assuming the number of offspring of each mated female is Poisson distributed, simulations show that the standard deviation of the fitness measure (the total number of offspring of the 157 tested females divided by the total number of offspring of the 157 reference females) was between 0.02 and 0.06 depending on the average of the Poisson distribution (allowed to vary between 3 to 20). We chose 0.04 as a rough estimate of the standard error of the productivity fitness measure. We were unable to perform a similar calculation for mating success, as the number of replicates was not reported for this assay.

| | | | | | | | | |
|---|---|---|---|---|---|---|---|---|
| 0 | 0 | 0 | 0 | 0 | 0.793 | 0.04 | 0 | 0.0506 |
| 1 | 0 | 0 | 0 | 0 | 0.427 | 0.04 | -0.618 | 0.0947 |
| 0 | 1 | 0 | 0 | 0 | 0.732 | 0.04 | -0.08 | 0.0549 |
| 0 | 0 | 1 | 0 | 0 | 0.807 | 0.04 | 0.018 | 0.0497 |
| 0 | 0 | 0 | 1 | 0 | 0.429 | 0.04 | -0.615 | 0.0944 |
| 0 | 0 | 0 | 0 | 1 | 1.66 | 0.04 | 0.739 | 0.0241 |
| 1 | 1 | 0 | 0 | 0 | 0.788 | 0.04 | -0.006 | 0.0509 |
| 1 | 0 | 1 | 0 | 0 | 0.613 | 0.04 | -0.258 | 0.0657 |
| 1 | 0 | 0 | 1 | 0 | 0.357 | 0.04 | -0.798 | 0.114 |
| 1 | 0 | 0 | 0 | 1 | 1.26 | 0.04 | 0.464 | 0.0318 |
| 0 | 1 | 1 | 0 | 0 | 0.38 | 0.04 | -0.736 | 0.107 |
| 0 | 1 | 0 | 1 | 0 | 0.262 | 0.04 | -1.11 | 0.157 |
| 0 | 1 | 0 | 0 | 1 | 0.967 | 0.04 | 0.198 | 0.0415 |
| 0 | 0 | 1 | 1 | 0 | 0.23 | 0.04 | -1.24 | 0.182 |
| 0 | 0 | 1 | 0 | 1 | 0.478 | 0.04 | -0.507 | 0.0846 |
| 0 | 0 | 0 | 1 | 1 | 1.24 | 0.04 | 0.45 | 0.0322 |
| 1 | 1 | 1 | 0 | 0 | 0.491 | 0.04 | -0.48 | 0.0823 |
| 1 | 1 | 0 | 1 | 0 | 0.162 | 0.04 | -1.59 | 0.271 |
| 1 | 1 | 0 | 0 | 1 | 0.589 | 0.04 | -0.297 | 0.0683 |
| 1 | 0 | 1 | 1 | 0 | 0.456 | 0.04 | -0.554 | 0.0887 |
| 1 | 0 | 1 | 0 | 1 | 0.823 | 0.04 | 0.037 | 0.0488 |
| 1 | 0 | 0 | 1 | 1 | 0.527 | 0.04 | -0.409 | 0.0765 |
| 0 | 1 | 1 | 1 | 0 | 0.143 | 0.04 | -1.71 | 0.316 |
| 0 | 1 | 1 | 0 | 1 | 0.954 | 0.04 | 0.185 | 0.042 |
| 0 | 1 | 0 | 1 | 1 | 1.03 | 0.04 | 0.258 | 0.039 |
| 0 | 0 | 1 | 1 | 1 | 0.274 | 0.04 | -1.06 | 0.151 |
| 1 | 1 | 1 | 1 | 0 | 0.0866 | 0.04 | -2.21 | 0.806 |
| 1 | 1 | 1 | 0 | 1 | 0.139 | 0.04 | -1.74 | 0.328 |
| 1 | 1 | 0 | 1 | 1 | 0.307 | 0.04 | -0.948 | 0.133 |
| 1 | 0 | 1 | 1 | 1 | 0.359 | 0.04 | -0.792 | 0.113 |
| 0 | 1 | 1 | 1 | 1 | 0.156 | 0.04 | -1.62 | 0.283 |
| 1 | 1 | 1 | 1 | 1 | 0.0105 | 0.04 | -4.33 | 1.13 |

## C2

Two genotypes had zero mating success and we subsequently excluded them from the analysis.

| | | | | | | | | |
|---|---|---|---|---|---|---|---|---|
| 0 | 0 | 0 | 0 | 0 | 1.82 | 0 | 0 | 0 |
| 1 | 0 | 0 | 0 | 0 | 4.72 | 0 | 0.953 | 0 |
| 0 | 1 | 0 | 0 | 0 | 1.2 | 0 | -0.416 | 0 |
| 0 | 0 | 1 | 0 | 0 | 1.13 | 0 | -0.48 | 0 |
| 0 | 0 | 0 | 1 | 0 | 0.444 | 0 | -1.41 | 0 |
| 0 | 0 | 0 | 0 | 1 | 1.6 | 0 | -0.128 | 0 |
| 1 | 1 | 0 | 0 | 0 | 0.791 | 0 | -0.832 | 0 |
| 1 | 0 | 1 | 0 | 0 | 3.42 | 0 | 0.631 | 0 |
| 1 | 0 | 0 | 1 | 0 | 0.47 | 0 | -1.35 | 0 |
| 1 | 0 | 0 | 0 | 1 | 1.17 | 0 | -0.444 | 0 |
| 0 | 1 | 1 | 0 | 0 | 0.0625 | 0 | -3.37 | 0 |
| 0 | 1 | 0 | 1 | 0 | 0.0612 | 0 | -3.39 | 0 |
| 0 | 1 | 0 | 0 | 1 | 1 | 0 | -0.598 | 0 |
| 0 | 0 | 1 | 1 | 0 | 0.211 | 0 | -2.16 | 0 |
| 0 | 0 | 1 | 0 | 1 | 0.444 | 0 | -1.41 | 0 |
| 0 | 0 | 0 | 1 | 1 | 0.357 | 0 | -1.63 | 0 |
| 1 | 1 | 1 | 0 | 0 | 0.316 | 0 | -1.75 | 0 |
| 1 | 1 | 0 | 1 | 0 | 0.0953 | 0 | -2.95 | 0 |
| 1 | 1 | 0 | 0 | 1 | 1.33 | 0 | -0.31 | 0 |
| 1 | 0 | 1 | 1 | 0 | 0.222 | 0 | -2.1 | 0 |
| 1 | 0 | 1 | 0 | 1 | 0.394 | 0 | -1.53 | 0 |
| 1 | 0 | 0 | 1 | 1 | 0.4 | 0 | -1.51 | 0 |
| 0 | 1 | 1 | 1 | 0 | 0.363 | 0 | -1.61 | 0 |
| 0 | 1 | 0 | 1 | 1 | 0.261 | 0 | -1.94 | 0 |
| 0 | 0 | 1 | 1 | 1 | 0.115 | 0 | -2.76 | 0 |
| 1 | 1 | 1 | 1 | 0 | 0.125 | 0 | -2.68 | 0 |
| 1 | 1 | 1 | 0 | 1 | 0.313 | 0 | -1.76 | 0 |
| 1 | 1 | 0 | 1 | 1 | 0.1 | 0 | -2.9 | 0 |
| 1 | 0 | 1 | 1 | 1 | 0.333 | 0 | -1.7 | 0 |
| 1 | 1 | 1 | 1 | 1 | 0.25 | 0 | -1.98 | 0 |

## D

Data is available in (Rokyta et al. 2011), Table 3. The fitness measure was the growth rate of the phage population on E. coli strain ($\log_2$ increase in the phage population, per hour). Standard errors were reported, and we converted these to standard errors on the log-scale using the procedure outline above. In this case, because the growth rate of the ancestor was reported, we could have computed the log-fitness as $(r_m - r_0)/r_0$. However, we used $\log[r_m/r_0]$ because both values were almost identical.

| | | | | | | | | | | | | |
|---|---|---|---|---|---|---|---|---|---|---|---|---|
| 0 | 0 | 0 | 0 | 0 | 0 | 0 | 0 | 0 | 15.2 | 0.2 | 0 | 0.0132 |
| 1 | 0 | 0 | 0 | 0 | 0 | 0 | 0 | 0 | 19.1 | 0.19 | 0.228 | 0.00996 |
| 0 | 1 | 0 | 0 | 0 | 0 | 0 | 0 | 0 | 19.3 | 0.43 | 0.242 | 0.0222 |
| 0 | 0 | 1 | 0 | 0 | 0 | 0 | 0 | 0 | 19.4 | 0.56 | 0.243 | 0.029 |
| 0 | 0 | 0 | 1 | 0 | 0 | 0 | 0 | 0 | 18.6 | 0.49 | 0.204 | 0.0263 |
| 0 | 0 | 0 | 0 | 1 | 0 | 0 | 0 | 0 | 16.8 | 0.36 | 0.104 | 0.0214 |
| 0 | 0 | 0 | 0 | 0 | 1 | 0 | 0 | 0 | 18.6 | 0.37 | 0.202 | 0.0199 |
| 0 | 0 | 0 | 0 | 0 | 0 | 1 | 0 | 0 | 21 | 0.26 | 0.325 | 0.0124 |
| 0 | 0 | 0 | 0 | 0 | 0 | 0 | 1 | 0 | 18.6 | 0.42 | 0.204 | 0.0226 |
| 0 | 0 | 0 | 0 | 0 | 0 | 0 | 0 | 1 | 16.6 | 0.28 | 0.0894 | 0.0169 |
| 1 | 1 | 0 | 0 | 0 | 0 | 0 | 0 | 0 | 18.7 | 0.25 | 0.206 | 0.0134 |
| 1 | 0 | 0 | 1 | 0 | 0 | 0 | 0 | 0 | 18.8 | 0.37 | 0.215 | 0.0197 |
| 1 | 0 | 0 | 0 | 0 | 1 | 0 | 0 | 0 | 18.1 | 0.56 | 0.174 | 0.031 |
| 1 | 0 | 0 | 0 | 0 | 1 | 0 | 0 | 0 | 22.5 | 0.25 | 0.394 | 0.0111 |
| 1 | 0 | 0 | 0 | 0 | 0 | 1 | 0 | 0 | 20.3 | 0.29 | 0.29 | 0.0143 |
| 0 | 1 | 0 | 1 | 0 | 0 | 0 | 0 | 0 | 17.8 | 0.33 | 0.158 | 0.0186 |
| 0 | 1 | 0 | 0 | 1 | 0 | 0 | 0 | 0 | 15.6 | 0.53 | 0.026 | 0.0341 |
| 0 | 1 | 0 | 0 | 0 | 0 | 1 | 0 | 0 | 17.3 | 0.54 | 0.131 | 0.0312 |
| 0 | 1 | 0 | 0 | 0 | 0 | 0 | 1 | 0 | 19.5 | 0.43 | 0.249 | 0.0221 |
| 0 | 1 | 0 | 0 | 0 | 0 | 0 | 0 | 1 | 16.5 | 0.48 | 0.0846 | 0.0291 |
| 0 | 0 | 1 | 1 | 0 | 0 | 0 | 0 | 0 | 17.5 | 0.36 | 0.143 | 0.0206 |
| 0 | 0 | 1 | 0 | 1 | 0 | 0 | 0 | 0 | 11.6 | 0.47 | -0.272 | 0.0407 |
| 0 | 0 | 1 | 0 | 0 | 0 | 1 | 0 | 0 | 19.5 | 0.28 | 0.25 | 0.0144 |
| 0 | 0 | 0 | 1 | 0 | 1 | 0 | 0 | 0 | 19.3 | 0.31 | 0.24 | 0.0161 |
| 0 | 0 | 0 | 1 | 0 | 0 | 0 | 1 | 0 | 18.5 | 0.43 | 0.197 | 0.0233 |
| 0 | 0 | 0 | 1 | 0 | 0 | 0 | 0 | 1 | 15.4 | 0.34 | 0.0144 | 0.0221 |
| 0 | 0 | 0 | 0 | 1 | 0 | 1 | 0 | 0 | 16.5 | 0.44 | 0.0858 | 0.0266 |
| 0 | 0 | 0 | 0 | 1 | 0 | 0 | 0 | 1 | 12.7 | 0.35 | -0.175 | 0.0275 |

**E1**

Data available in (Sanjuán et al. 2004), supplementary table 1. The fitness measure was the growth rate of the mutant relative to the ancestral strain. The absolute growth rate of the ancestor was not reported. Standard errors were reported, and we converted them to standard errors on the log-scale using the procedure outlined above.

| | | | | | | | | | |
|---|---|---|---|---|---|---|---|---|---|
| 0 | 0 | 0 | 0 | 0 | 0 | 1 | 0 | 0 | 0 |
| 1 | 0 | 0 | 0 | 0 | 0 | 1.01 | 0.015 | 0.0129 | 0.0148 |
| 0 | 1 | 0 | 0 | 0 | 0 | 1 | 0.014 | 0 | 0.014 |
| 0 | 0 | 1 | 0 | 0 | 0 | 1.01 | 0.025 | 0.0139 | 0.0247 |
| 0 | 0 | 0 | 1 | 0 | 0 | 1.03 | 0.016 | 0.0296 | 0.0155 |
| 0 | 0 | 0 | 0 | 1 | 0 | 1.1 | 0.022 | 0.0953 | 0.02 |
| 0 | 0 | 0 | 0 | 0 | 1 | 1.09 | 0.028 | 0.088 | 0.0257 |
| 1 | 1 | 0 | 0 | 0 | 0 | 0.936 | 0.016 | -0.0661 | 0.0171 |
| 1 | 0 | 1 | 0 | 0 | 0 | 0.885 | 0.016 | -0.122 | 0.0181 |
| 1 | 0 | 0 | 1 | 0 | 0 | 0.92 | 0.012 | -0.0834 | 0.013 |
| 1 | 0 | 0 | 0 | 1 | 0 | 1.03 | 0.015 | 0.0257 | 0.0146 |
| 1 | 0 | 0 | 0 | 0 | 1 | 0.978 | 0.009 | -0.0222 | 0.0092 |
| 0 | 1 | 1 | 0 | 0 | 0 | 0.93 | 0.013 | -0.0726 | 0.014 |
| 0 | 1 | 0 | 1 | 0 | 0 | 0.998 | 0.013 | -0.002 | 0.013 |
| 0 | 1 | 0 | 0 | 1 | 0 | 0.942 | 0.014 | -0.0598 | 0.0149 |
| 0 | 1 | 0 | 0 | 0 | 1 | 1.06 | 0.02 | 0.0611 | 0.0188 |
| 0 | 0 | 1 | 1 | 0 | 0 | 1.06 | 0.017 | 0.062 | 0.016 |
| 0 | 0 | 1 | 0 | 1 | 0 | 1.09 | 0.031 | 0.0871 | 0.0284 |
| 0 | 0 | 1 | 0 | 0 | 1 | 1.08 | 0.024 | 0.0788 | 0.0222 |
| 0 | 0 | 0 | 1 | 1 | 0 | 0.928 | 0.012 | -0.0747 | 0.0129 |
| 0 | 0 | 0 | 1 | 0 | 1 | 1.11 | 0.031 | 0.106 | 0.0279 |
| 0 | 0 | 0 | 0 | 1 | 1 | 1.12 | 0.026 | 0.11 | 0.0233 |

**E2**

Data available in (Sanjuán et al. 2004), supplementary table 1. Fitness measure and standard errors as for E1. The table, including 28 mutation and 76 fitness values, is too large to show here, but is available upon request.

**F**

Data available in (Khan et al. 2011), supplementary table S2. Fitness was the growth rate of the strain relative to that of the ancestor, measured in a direct competition assay. The absolute growth rate of the ancestor was not reported. 95% confidence intervals were reported, and we used this information to get an approximate standard error on the fitness scale and on the log-fitness scale.

| | | | | | | | | |
|---|---|---|---|---|---|---|---|---|
| 0 | 0 | 0 | 0 | 0 | 0.997 | 0.005  | 0      | 0.00502 |
| 1 | 0 | 0 | 0 | 0 | 1.01  | 0.0065 | 0.0149 | 0.00642 |
| 0 | 1 | 0 | 0 | 0 | 1.14  | 0.0115 | 0.136  | 0.0101  |
| 0 | 0 | 1 | 0 | 0 | 1.1   | 0.0085 | 0.103  | 0.00769 |
| 0 | 0 | 0 | 1 | 0 | 1.03  | 0.008  | 0.0296 | 0.00779 |
| 0 | 0 | 0 | 0 | 1 | 1     | 0.0065 | 0.003  | 0.0065  |
| 1 | 1 | 0 | 0 | 0 | 1.1   | 0.009  | 0.101  | 0.00816 |
| 1 | 0 | 1 | 0 | 0 | 1.12  | 0.0095 | 0.115  | 0.0085  |
| 1 | 0 | 0 | 1 | 0 | 1.05  | 0.0065 | 0.0489 | 0.00621 |
| 1 | 0 | 0 | 0 | 1 | 1.02  | 0.009  | 0.0248 | 0.00881 |
| 0 | 1 | 1 | 0 | 0 | 1.2   | 0.0105 | 0.189  | 0.00871 |
| 0 | 1 | 0 | 1 | 0 | 1.15  | 0.0075 | 0.142  | 0.00653 |
| 0 | 1 | 0 | 0 | 1 | 1.19  | 0.0115 | 0.179  | 0.00964 |
| 0 | 0 | 1 | 1 | 0 | 1.12  | 0.006  | 0.12   | 0.00534 |
| 0 | 0 | 1 | 0 | 1 | 1.18  | 0.0115 | 0.173  | 0.00971 |
| 0 | 0 | 0 | 1 | 1 | 1.08  | 0.0135 | 0.0753 | 0.0126  |
| 1 | 1 | 1 | 0 | 0 | 1.2   | 0.0105 | 0.183  | 0.00877 |
| 1 | 1 | 0 | 1 | 0 | 1.16  | 0.0125 | 0.15   | 0.0108  |
| 1 | 1 | 0 | 0 | 1 | 1.19  | 0.012  | 0.176  | 0.0101  |
| 1 | 0 | 1 | 1 | 0 | 1.12  | 0.0135 | 0.12   | 0.012   |
| 1 | 0 | 1 | 0 | 1 | 1.2   | 0.0105 | 0.189  | 0.00872 |
| 1 | 0 | 0 | 1 | 1 | 1.13  | 0.01   | 0.122  | 0.00888 |
| 0 | 1 | 1 | 1 | 0 | 1.2   | 0.0155 | 0.186  | 0.0129  |
| 0 | 1 | 1 | 0 | 1 | 1.28  | 0.014  | 0.25   | 0.0109  |
| 0 | 1 | 0 | 1 | 1 | 1.22  | 0.0195 | 0.199  | 0.016   |
| 0 | 0 | 1 | 1 | 1 | 1.19  | 0.012  | 0.174  | 0.0101  |
| 1 | 1 | 1 | 1 | 0 | 1.23  | 0.0105 | 0.208  | 0.00855 |
| 1 | 1 | 1 | 0 | 1 | 1.28  | 0.0195 | 0.248  | 0.0153  |
| 1 | 1 | 0 | 1 | 1 | 1.19  | 0.015  | 0.177  | 0.0126  |
| 1 | 0 | 1 | 1 | 1 | 1.21  | 0.008  | 0.195  | 0.0066  |
| 0 | 1 | 1 | 1 | 1 | 1.3   | 0.0155 | 0.265  | 0.0119  |
| 1 | 1 | 1 | 1 | 1 | 1.32  | 0.0155 | 0.281  | 0.0117  |

## G

Data available in (Chou et al. 2011), figure 1. The fitness measure was the growth rate relative to the ancestral strain, measured in a competition assay. The absolute growth rate of the ancestor was not reported. Standard errors for the fitness of genotypes in the fitness landscape were not reported. However, the standard errors of fitness measurement were reported for a set of strains isolated at several generations and in several replicates of the evolution experiment (supplementary table 1), allowing us to estimate the magnitude of error for fitness measurements done in this study. There was a significant positive correlation between the fitness value and its standard error. We described this correlation with a linear model, which we then used to compute the predicted standard error for each fitness value in the fitness landscape.

| | | | | | | | |
|---|---|---|---|---|---|---|---|
| 0 | 0 | 0 | 0 | 1    | 0.0203 | 0      | 0.0203 |
| 0 | 0 | 0 | 1 | 1.17 | 0.0247 | 0.154  | 0.0212 |
| 0 | 1 | 0 | 0 | 1.1  | 0.0228 | 0.0917 | 0.0208 |
| 1 | 0 | 0 | 0 | 1.14 | 0.024  | 0.133  | 0.0211 |
| 0 | 0 | 1 | 0 | 1.51 | 0.0338 | 0.411  | 0.0224 |
| 1 | 0 | 1 | 0 | 1.62 | 0.0368 | 0.484  | 0.0227 |
| 0 | 1 | 1 | 0 | 1.61 | 0.0366 | 0.479  | 0.0227 |
| 1 | 1 | 0 | 0 | 1.28 | 0.0277 | 0.248  | 0.0217 |
| 0 | 0 | 1 | 1 | 1.64 | 0.0372 | 0.494  | 0.0227 |
| 1 | 0 | 0 | 1 | 1.32 | 0.0288 | 0.278  | 0.0218 |
| 0 | 1 | 0 | 1 | 1.3  | 0.0282 | 0.262  | 0.0217 |
| 1 | 1 | 1 | 0 | 1.75 | 0.0402 | 0.561  | 0.023  |
| 1 | 0 | 1 | 1 | 1.78 | 0.0411 | 0.579  | 0.023  |
| 0 | 1 | 1 | 1 | 1.81 | 0.0418 | 0.594  | 0.0231 |
| 1 | 1 | 0 | 1 | 1.44 | 0.0318 | 0.361  | 0.0222 |
| 1 | 1 | 1 | 1 | 1.94 | 0.0451 | 0.66   | 0.0233 |

## H1

The fitness measure was resistance to cefotaxime quantified by Minimum Inhibitory Concentration expressed in µg/mL and is available in (Weinreich et al. 2006), supplementary information, table S1. Three replicate measurements were done, from which we could calculate the average MIC and standard errors. Standard errors were converted to the log scale using the procedure outline above. Note that in that case standard errors were often 0, as they did not include the error due to the fact that the assay can only give a discrete number of MIC values.

| | | | | | | | | |
|---|---|---|---|---|---|---|---|---|
| 0 | 0 | 0 | 0 | 0 | 0.088 | 0 | 0 | 0 |
| 0 | 0 | 0 | 0 | 1 | 1.4 | 0 | 2.77 | 0 |
| 0 | 0 | 0 | 1 | 0 | 0.063 | 0.0144 | -0.334 | 0.248 |
| 0 | 0 | 0 | 1 | 1 | 32 | 0 | 5.9 | 0 |
| 0 | 0 | 1 | 0 | 0 | 0.13 | 0.0461 | 0.39 | 0.442 |
| 0 | 0 | 1 | 0 | 1 | 360 | 0 | 8.32 | 0 |
| 0 | 0 | 1 | 1 | 0 | 0.18 | 0.0289 | 0.716 | 0.166 |
| 0 | 0 | 1 | 1 | 1 | 360 | 0 | 8.32 | 0 |
| 0 | 1 | 0 | 0 | 0 | 0.088 | 0 | 0 | 0 |
| 0 | 1 | 0 | 0 | 1 | 23 | 0 | 5.57 | 0 |
| 0 | 1 | 0 | 1 | 0 | 1.4 | 0 | 2.77 | 0 |
| 0 | 1 | 0 | 1 | 1 | 360 | 0 | 8.32 | 0 |
| 0 | 1 | 1 | 0 | 0 | 1.4 | 0 | 2.77 | 0 |
| 0 | 1 | 1 | 0 | 1 | 2100 | 0 | 10.1 | 0 |
| 0 | 1 | 1 | 1 | 0 | 0.71 | 0.167 | 2.09 | 0.256 |
| 0 | 1 | 1 | 1 | 1 | 2900 | 0 | 10.4 | 0 |
| 1 | 0 | 0 | 0 | 0 | 0.088 | 0 | 0 | 0 |
| 1 | 0 | 0 | 0 | 1 | 1.4 | 0 | 2.77 | 0 |
| 1 | 0 | 0 | 1 | 0 | 0.088 | 0 | 0 | 0 |
| 1 | 0 | 0 | 1 | 1 | 360 | 0 | 8.32 | 0 |
| 1 | 0 | 1 | 0 | 0 | 0.18 | 0 | 0.716 | 0 |
| 1 | 0 | 1 | 0 | 1 | 360 | 0 | 8.32 | 0 |
| 1 | 0 | 1 | 1 | 0 | 0.18 | 0 | 0.716 | 0 |
| 1 | 0 | 1 | 1 | 1 | 2100 | 0 | 10.1 | 0 |
| 1 | 1 | 0 | 0 | 0 | 0.088 | 0 | 0 | 0 |
| 1 | 1 | 0 | 0 | 1 | 360 | 0 | 8.32 | 0 |
| 1 | 1 | 0 | 1 | 0 | 0.088 | 0 | 0 | 0 |
| 1 | 1 | 0 | 1 | 1 | 360 | 0 | 8.32 | 0 |
| 1 | 1 | 1 | 0 | 0 | 2 | 0.702 | 3.12 | 0.436 |
| 1 | 1 | 1 | 0 | 1 | 1500 | 346 | 9.74 | 0.25 |
| 1 | 1 | 1 | 1 | 0 | 1.4 | 0 | 2.77 | 0 |
| 1 | 1 | 1 | 1 | 1 | 4100 | 0 | 10.7 | 0 |

**H2**

The fitness measure is mean resistance to cefotaxime quantified by Minimum Inhibitory Concentration expressed in μg/mL. and is available in (Tan et al. 2011), supplementary information, table 1 (reported in logarithm base $\sqrt{2}$). The standard errors of MIC were reported; in that dataset, the error included both measurement error and the error due to the fact that the assay can only give a discrete number of MIC values. Because errors were reported on the log scale (base $\sqrt{2}$), we directly calculated the error on our (natural) log scale.

| | | | | | | | | |
|---|---|---|---|---|---|---|---|---|
| 0 | 0 | 0 | 0 | 0 | 0.0625 | - | 0 | 0.173 |
| 0 | 0 | 0 | 0 | 1 | 0.793 | - | 2.54 | 0.208 |
| 0 | 0 | 0 | 1 | 0 | 0.0701 | - | 0.114 | 0.208 |
| 0 | 0 | 0 | 1 | 1 | 80.7 | - | 7.16 | 0.208 |
| 0 | 0 | 1 | 0 | 0 | 0.125 | - | 0.693 | 0.173 |
| 0 | 0 | 1 | 0 | 1 | 161 | - | 7.86 | 0.208 |
| 0 | 0 | 1 | 1 | 0 | 0.158 | - | 0.925 | 0.208 |
| 0 | 0 | 1 | 1 | 1 | 1020 | - | 9.7 | 0.173 |
| 0 | 1 | 0 | 0 | 0 | 0.111 | - | 0.579 | 0.208 |
| 0 | 1 | 0 | 0 | 1 | 101 | - | 7.39 | 0.208 |
| 0 | 1 | 0 | 1 | 0 | 0.0788 | - | 0.232 | 0.208 |
| 0 | 1 | 0 | 1 | 1 | 181 | - | 7.97 | 0.173 |
| 0 | 1 | 1 | 0 | 0 | 1.12 | - | 2.89 | 0.208 |
| 0 | 1 | 1 | 0 | 1 | 1020 | - | 9.7 | 0.173 |
| 0 | 1 | 1 | 1 | 0 | 0.631 | - | 2.31 | 0.208 |
| 0 | 1 | 1 | 1 | 1 | 1450 | - | 10.1 | 0.173 |
| 1 | 0 | 0 | 0 | 0 | 0.0991 | - | 0.461 | 0.208 |
| 1 | 0 | 0 | 0 | 1 | 1.12 | - | 2.89 | 0.208 |
| 1 | 0 | 0 | 1 | 0 | 0.0701 | - | 0.114 | 0.208 |
| 1 | 0 | 0 | 1 | 1 | 128 | - | 7.62 | 0.173 |
| 1 | 0 | 1 | 0 | 0 | 0.25 | - | 1.39 | 0.173 |
| 1 | 0 | 1 | 0 | 1 | 181 | - | 7.97 | 0.173 |
| 1 | 0 | 1 | 1 | 0 | 0.198 | - | 1.15 | 0.208 |
| 1 | 0 | 1 | 1 | 1 | 1020 | - | 9.7 | 0.173 |
| 1 | 1 | 0 | 0 | 0 | 0.0701 | - | 0.114 | 0.208 |
| 1 | 1 | 0 | 0 | 1 | 181 | - | 7.97 | 0.173 |
| 1 | 1 | 0 | 1 | 0 | 0.0701 | - | 0.114 | 0.208 |
| 1 | 1 | 0 | 1 | 1 | 203 | - | 8.09 | 0.208 |
| 1 | 1 | 1 | 0 | 0 | 1.78 | - | 3.35 | 0.208 |
| 1 | 1 | 1 | 0 | 1 | 1150 | - | 9.82 | 0.208 |
| 1 | 1 | 1 | 1 | 0 | 1.78 | - | 3.35 | 0.208 |
| 1 | 1 | 1 | 1 | 1 | 2050 | - | 10.4 | 0.173 |

**H3**

For datasets H3 and H4 the fitness measure was cefotaxime resistance, measured as IC99,99. The dataset is available in (Schenk et al. 2013) supplementary information, supplementary table 1 (H4, large effect mutations) and supplementary table 2 (H3, small effect mutations). Standard errors were not reported, but because the assay is very similar to that used for H1 and H2 landscapes, we chose an error of the same order of magnitude as the one in H2, $\sigma_{error}^{log} = 0.2$.

| | | | | | | | |
|---|---|---|---|---|---|---|---|
| 0 | 0 | 0 | 0 | 0.053 | - | 0 | 0.2 |
| 1 | 0 | 0 | 1 | 0.045 | - | -0.164 | 0.2 |
| 1 | 0 | 1 | 1 | 0.05 | - | -0.0583 | 0.2 |
| 1 | 1 | 1 | 1 | 0.063 | - | 0.173 | 0.2 |
| 1 | 1 | 0 | 1 | 0.069 | - | 0.264 | 0.2 |
| 0 | 1 | 0 | 0 | 0.069 | - | 0.264 | 0.2 |
| 1 | 0 | 0 | 0 | 0.077 | - | 0.374 | 0.2 |
| 1 | 1 | 0 | 0 | 0.08 | - | 0.412 | 0.2 |
| 0 | 0 | 1 | 1 | 0.086 | - | 0.484 | 0.2 |
| 1 | 0 | 1 | 0 | 0.092 | - | 0.551 | 0.2 |
| 0 | 0 | 0 | 1 | 0.093 | - | 0.562 | 0.2 |
| 0 | 0 | 1 | 0 | 0.104 | - | 0.674 | 0.2 |
| 1 | 1 | 1 | 0 | 0.118 | - | 0.8 | 0.2 |
| 0 | 1 | 0 | 1 | 0.123 | - | 0.842 | 0.2 |
| 0 | 1 | 1 | 1 | 0.138 | - | 0.957 | 0.2 |
| 0 | 1 | 1 | 0 | 0.163 | - | 1.12 | 0.2 |

**H4**

| | | | | | | | |
|---|---|---|---|---|---|---|---|
| 0 | 0 | 0 | 0 | 0.053 | - | 0 | 0.2 |
| 0 | 1 | 0 | 1 | 0.099 | - | 0.625 | 0.2 |
| 1 | 1 | 1 | 1 | 0.128 | - | 0.882 | 0.2 |
| 1 | 0 | 0 | 0 | 0.183 | - | 1.24 | 0.2 |
| 0 | 1 | 1 | 0 | 0.205 | - | 1.35 | 0.2 |
| 1 | 1 | 0 | 1 | 0.217 | - | 1.41 | 0.2 |
| 0 | 1 | 1 | 1 | 0.28 | - | 1.66 | 0.2 |
| 0 | 1 | 0 | 0 | 0.424 | - | 2.08 | 0.2 |
| 0 | 0 | 0 | 1 | 0.441 | - | 2.12 | 0.2 |
| 0 | 0 | 1 | 1 | 0.886 | - | 2.82 | 0.2 |
| 1 | 0 | 0 | 1 | 1.2 | - | 3.12 | 0.2 |
| 0 | 0 | 1 | 0 | 1.22 | - | 3.14 | 0.2 |
| 1 | 1 | 1 | 0 | 1.88 | - | 3.57 | 0.2 |
| 1 | 0 | 1 | 1 | 1.91 | - | 3.58 | 0.2 |
| 1 | 1 | 0 | 0 | 4.01 | - | 4.33 | 0.2 |
| 1 | 0 | 1 | 0 | 11.7 | - | 5.4 | 0.2 |

**I1**

In this dataset the fitness measure was pyrimethamine resistance, measured as $IC_{50}$. $IC_{50}$ is the pyrimethamine concentration (expressed in μg/mL) at which the strain's growth rate is 50% that achieved in the absence of pyrimethamine and correlated very well with Minimum Inhibitory Concentration. The dataset is available in (Lozovsky et al. 2009), supplementary information, supplementary table 1. Standard deviations of $IC_{50}$ were reported.

| | | | | | | | |
|---|---|---|---|---|---|---|---|
| 0 | 0 | 0 | 0 | 0.27 | 0.04 | 0 | 0.153 |
| 1 | 0 | 0 | 0 | 0.36 | 0.03 | 0.288 | 0.0841 |
| 0 | 1 | 0 | 0 | 2.04 | 0.65 | 2.02 | 0.377 |
| 0 | 0 | 1 | 0 | 9.56 | 0.95 | 3.57 | 0.101 |
| 0 | 0 | 0 | 1 | 0.29 | 0.05 | 0.0715 | 0.18 |
| 1 | 1 | 0 | 0 | 4.89 | 0.8 | 2.9 | 0.17 |
| 1 | 0 | 1 | 0 | 37.7 | 0.58 | 4.94 | 0.0154 |
| 1 | 0 | 0 | 1 | 103 | 1.53 | 5.94 | 0.0149 |
| 0 | 1 | 1 | 0 | 147 | 3.79 | 6.3 | 0.0258 |
| 0 | 1 | 0 | 1 | 7.28 | 0.37 | 3.29 | 0.051 |
| 0 | 0 | 1 | 1 | 0 | 0 | -Inf | - |
| 1 | 1 | 1 | 0 | 242 | 8.47 | 6.8 | 0.035 |
| 1 | 1 | 0 | 1 | 56 | 1.73 | 5.33 | 0.0309 |
| 1 | 0 | 1 | 1 | 56.8 | 0.77 | 5.35 | 0.0136 |
| 0 | 1 | 1 | 1 | 195 | 10 | 6.58 | 0.0517 |
| 1 | 1 | 1 | 1 | 300 | 7.46 | 7.01 | 0.0249 |

**I2**

In this dataset the fitness measure was pyrimethamine resistance, measured as $IC_{50}$, here expressed in mol/L. The dataset is available in (Brown et al. 2010), supplementary information, supplementary table 3 (the $\log_{10}(IC_{50})$ is reported). Because our framework includes only diallelic loci, we removed genotypes with the third allele. Standard errors were reported on the $\log_{10}$ scale, so we directly converted them to standard errors on our (natural) log scale.

| | | | | | | | | |
|---|---|---|---|---|---|---|---|---|
| 0 | 0 | 0 | 0 | 0 | 5.17e-07 | - | 0 | 0.121 |
| 0 | 0 | 0 | 0 | 1 | 1.54e-06 | - | 1.09 | 0.0302 |
| 0 | 0 | 0 | 1 | 0 | 5.77e-05 | - | 4.71 | 0.0322 |
| 0 | 0 | 1 | 0 | 0 | 9e-07 | - | 0.553 | 0.0813 |
| 0 | 0 | 1 | 0 | 1 | 1.68e-06 | - | 1.18 | 0.0442 |
| 0 | 0 | 1 | 1 | 0 | 0.000185 | - | 5.88 | 0.0566 |
| 0 | 0 | 1 | 1 | 1 | 0.000282 | - | 6.3 | 0.0762 |
| 0 | 1 | 0 | 0 | 0 | 1.89e-06 | - | 1.3 | 0.0659 |
| 0 | 1 | 0 | 0 | 1 | 3.23e-06 | - | 1.83 | 0.0691 |
| 0 | 1 | 0 | 1 | 0 | 9.66e-05 | - | 5.23 | 0.0396 |
| 0 | 1 | 0 | 1 | 1 | 2.51e-05 | - | 3.88 | 0.0762 |
| 0 | 1 | 1 | 0 | 0 | 1.69e-06 | - | 1.18 | 0.0636 |
| 0 | 1 | 1 | 0 | 1 | 2.38e-06 | - | 1.52 | 0.0792 |
| 0 | 1 | 1 | 1 | 0 | 0.000259 | - | 6.21 | 0.267 |
| 0 | 1 | 1 | 1 | 1 | 0.000501 | - | 6.88 | 0.0762 |
| 1 | 0 | 1 | 0 | 1 | 2.15e-06 | - | 1.42 | 0.0739 |
| 1 | 1 | 1 | 0 | 0 | 6.74e-07 | - | 0.265 | 0.177 |

**I3**

This dataset included mutations conferring pyrimethamine resistance, analogous to those studied in I1 and I2. The measure of fitness was the growth rate of a yeast transformed with *Plasmodium vivax* DHFR gene, in the presence of 1 μmol/L of pyrimethamine (Jiang et al. 2013). Data was extracted from fig. 2, and the identity of genotypes was deduced from their IC50 presented in fig. 1. We were unable to compute standard errors for these fitness measures. Here, because the absolute value of $r_0$ was given, we computed fitness as $(r_m - r_0)/r_0$. Indeed, selection was sufficiently strong in that case that using $\log[r_m/r_0]$ as an approximation would give an incorrect fitness measure.

| | | | | | | | |
|---|---|---|---|---|---|---|---|
| 0 | 0 | 0 | 0 | 0.0122 | - | 0 | - |
| 0 | 0 | 1 | 1 | 0.0219 | - | 0.788 | - |
| 1 | 0 | 1 | 1 | 0.0267 | - | 1.18 | - |
| 0 | 0 | 0 | 1 | 0.0306 | - | 1.5 | - |
| 0 | 1 | 0 | 0 | 0.0329 | - | 1.69 | - |
| 1 | 1 | 0 | 0 | 0.035 | - | 1.86 | - |
| 1 | 1 | 1 | 1 | 0.0371 | - | 2.03 | - |
| 0 | 0 | 1 | 0 | 0.0371 | - | 2.03 | - |
| 0 | 1 | 0 | 1 | 0.0376 | - | 2.08 | - |
| 1 | 0 | 1 | 0 | 0.0381 | - | 2.11 | - |
| 1 | 0 | 0 | 0 | 0.0386 | - | 2.16 | - |
| 0 | 1 | 1 | 1 | 0.0392 | - | 2.2 | - |
| 1 | 0 | 0 | 1 | 0.0392 | - | 2.2 | - |
| 0 | 1 | 1 | 0 | 0.0396 | - | 2.24 | - |
| 1 | 1 | 0 | 1 | 0.0401 | - | 2.28 | - |
| 1 | 1 | 1 | 0 | 0.0403 | - | 2.3 | - |

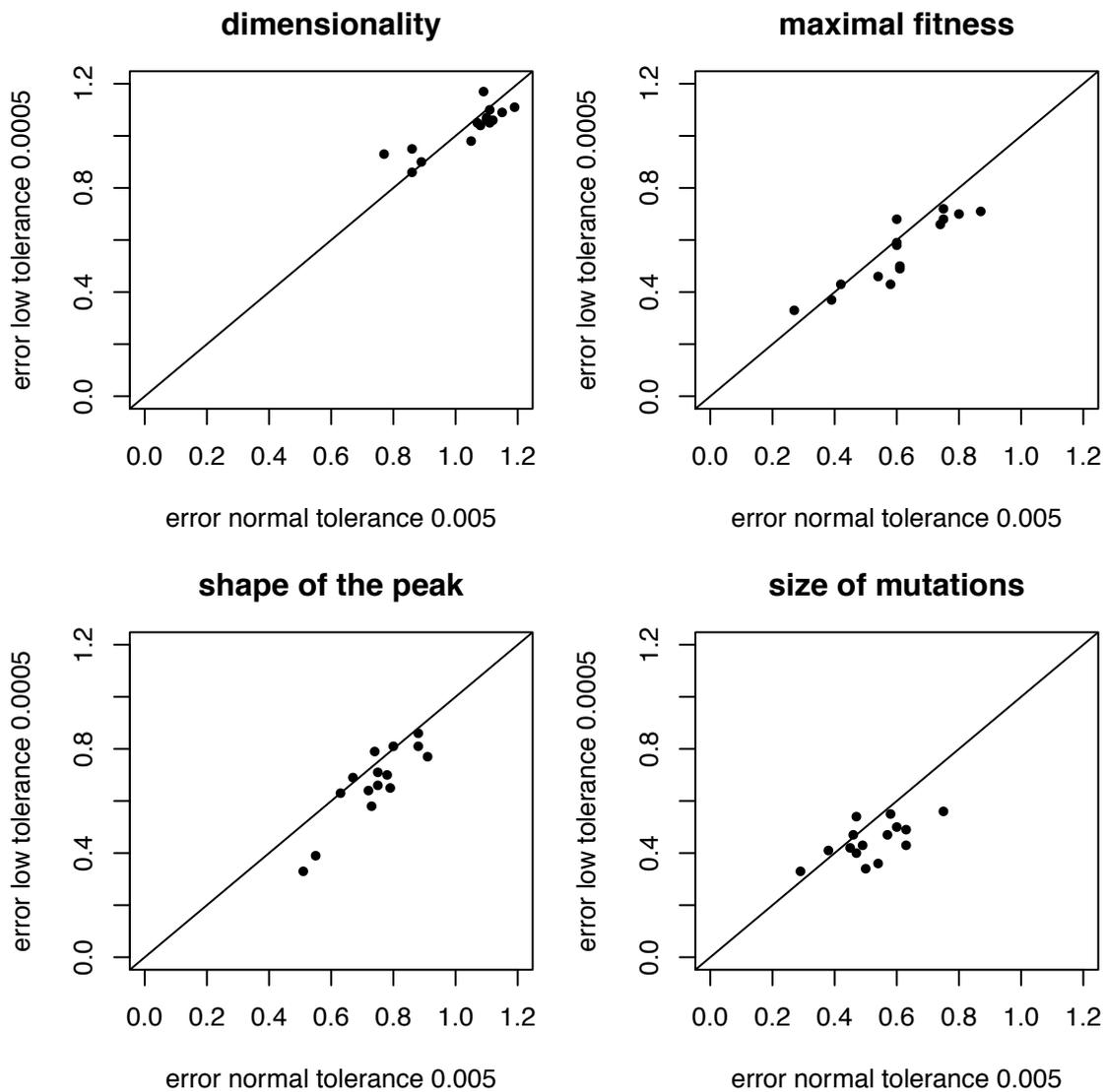

**Supplementary figure 1:** Comparison of the prediction errors for each parameter, when all fitness values are used, for tolerance equal to 0.005 (used throughout the study) versus a 10 times smaller tolerance equal to 0.0005. Each point is one of the protocols shown in Table S2. Lower tolerance does not generally allow much more precise inference.

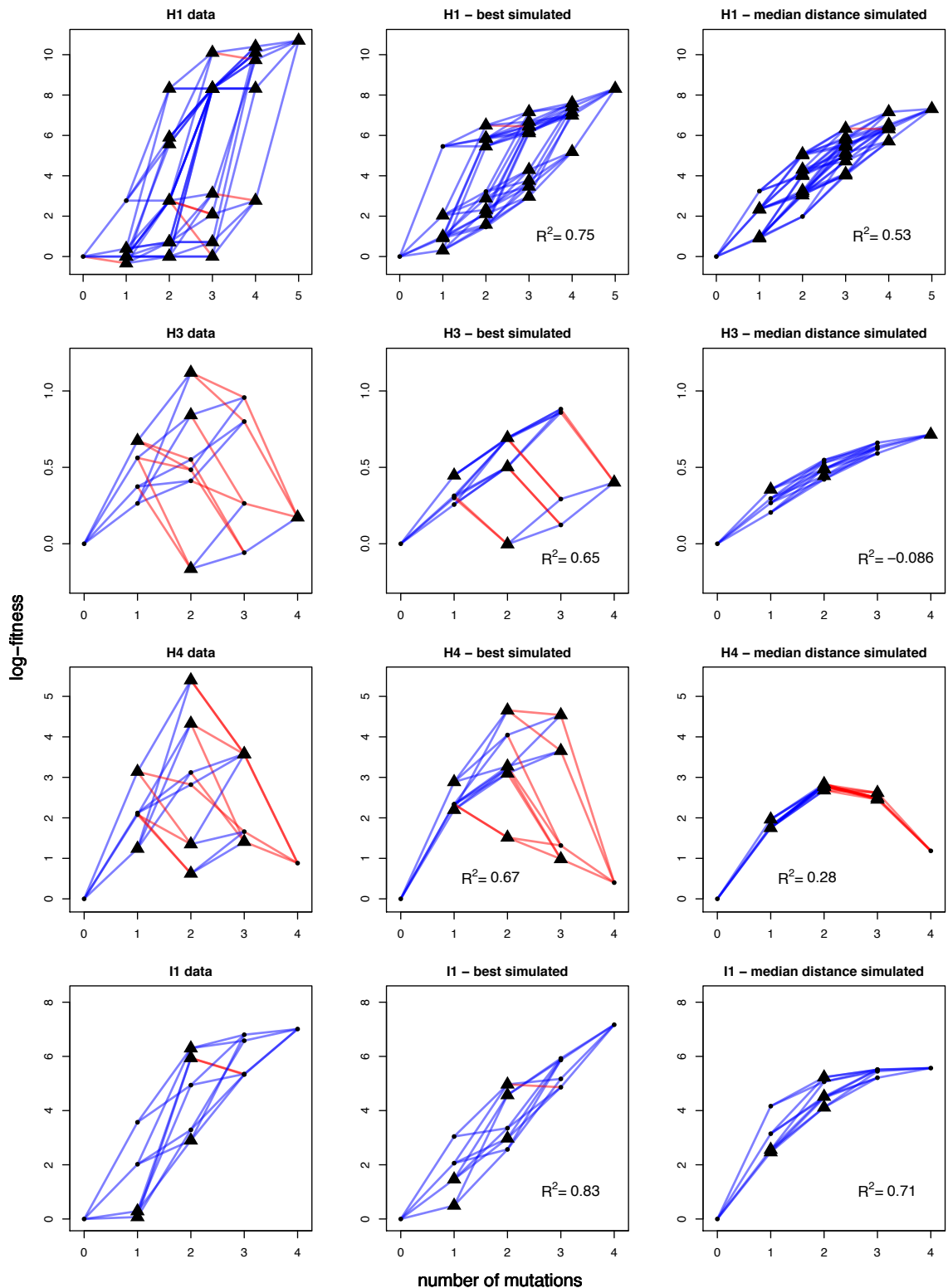

**Supplementary figure 2:** Empirical landscapes in the datasets of resistance to cefotaxime H1, H3 and H4, and resistance to pyrimethamine I1. For each dataset, the data (left) is shown side by side with the simulated genotypic landscape closest to the data in terms of Euclidean distance (middle), and a typical simulated landscape, defined as the landscape, among all simulated landscapes retained by the ABC framework, whose distance

to the data was closest to the median distance. Blue edges are beneficial mutations, red edges are deleterious mutations. Fitness values that are particularly unexpected under Fisher's model were marked with a triangle. Landscapes of drug resistance (H and I) strongly rejected Fisher's model ($p < 0.001$), except the landscape of cefotaxime resistance made of mutations of small effects (H3, $p = 0.07$). The fact that mutations have relatively small effect on fitness in the ancestral strain, but together confer extremely high fitness, is unexpected under Fisher's model. Sign epistasis with large effect mutations and far away from the optimum is also unexpected under Fisher's model.

**Table S1. Summary statistics for each dataset** (estimation ± standard deviation of error, when available)

| Name | Mean selection | Mean epistasis | Std dev selection | Std dev epistasis | Correlation fitness-epistasis | Maximal fitness |
|---|---|---|---|---|---|---|
| A1 | -0.188 ± 0.036 | 0 ± 0.019 | 0.056 ± 0.03 | 0.159 ± 0.042 | -0.39 ± 0.129 | 0 ± 0.058 |
| A2 | -0.161 ± 0.038 | -0.013 ± 0.02 | 0.062 ± 0.038 | 0.091 ± 0.033 | 0 ± 0.138 | 0 ± 0.062 |
| B1 | -0.078 ± 0.003 | 0.01 ± 0.011 | 0.182 ± 0.006 | 0.037 ± 0.007 | 0.02 ± 0.151 | 0.173 ± 0.048 |
| B2 | -0.048 ± 0.003 | -0.011 ± 0.009 | 0.146 ± 0.004 | 0.055 ± 0.008 | 0.189 ± 0.12 | 0.127 ± 0.083 |
| B3 | -0.042 ± 0.005 | 0.002 ± 0.012 | 0.142 ± 0.01 | 0.042 ± 0.012 | -0.106 ± 0.271 | 0.167 ± 0.016 |
| B4 | -0.152 ± 0.005 | -0.001 ± 0.012 | 0.265 ± 0.01 | 0.066 ± 0.013 | -0.224 ± 0.153 | 0.108 ± 0.048 |
| B5 | -0.021 ± 0.003 | 0 ± 0.008 | 0.115 ± 0.007 | 0.036 ± 0.01 | -0.09 ± 0.184 | 0.187 ± 0.034 |
| B6 | -0.042 ± 0.003 | -0.003 ± 0.01 | 0.155 ± 0.007 | 0.042 ± 0.009 | -0.2 ± 0.18 | 0.156 ± 0.033 |
| B7 | -0.099 ± 0.003 | -0.004 ± 0.009 | 0.236 ± 0.009 | 0.06 ± 0.008 | -0.145 ± 0.189 | 0.112 ± 0.029 |
| B8 | -0.13 ± 0.01 | -0.008 ± 0.015 | 0.356 ± 0.025 | 0.049 ± 0.012 | -0.105 ± 0.35 | 0.197 ± 0.023 |
| B9 | -0.031 ± 0.003 | 0.001 ± 0.01 | 0.134 ± 0.008 | 0.04 ± 0.014 | -0.134 ± 0.11 | 0.107 ± 0.063 |
| B10 | 0.001 ± 0.002 | -0.005 ± 0.009 | 0.068 ± 0.003 | 0.033 ± 0.007 | 0.087 ± 0.137 | 0.146 ± 0.042 |
| C1 | -0.111 ± 0.031 | -0.509 ± 0.134 | 0.559 ± 0.032 | 0.999 ± 0.177 | 0.159 ± 0.193 | 0.739 ± 0.025 |
| C2 | -0.296 | 0.271 | 0.848 | 1.37 | -0.554 | 0.953 |
| D | 0.205 ± 0.007 | -0.272 ± 0.022 | 0.072 ± 0.005 | 0.114 ± 0.011 | 0.215 ± 0.102 | 0.394 ± 0.01 |
| E1 | 0.04 ± 0.007 | -0.078 ± 0.016 | 0.041 ± 0.008 | 0.063 ± 0.008 | -0.027 ± 0.172 | 0.11 ± 0.019 |
| E2 | -0.11 ± 0.006 | 0.03 ± 0.014 | 0.137 ± 0.007 | 0.088 ± 0.01 | 0.153 ± 0.121 | 0.058 ± 0.039 |
| F | 0.057 ± 0.004 | -0.003 ± 0.002 | 0.059 ± 0.005 | 0.036 ± 0.004 | -0.128 ± 0.074 | 0.281 ± 0.01 |
| G | 0.197 ± 0.01 | -0.022 ± 0.009 | 0.145 ± 0.013 | 0.03 ± 0.013 | 0.124 ± 0.24 | 0.66 ± 0.024 |
| H1 | 0.565 ± 0.093 | 0.262 ± 0.018 | 1.259 ± 0.046 | 2.173 ± 0.049 | -0.423 ± 0.005 | 10.7 ± 0 |
| H2 | 0.877 ± 0.088 | 0.216 ± 0.039 | 0.954 ± 0.097 | 1.778 ± 0.08 | -0.47 ± 0.025 | 10.4 ± 0.153 |
| H3 | 0.468 ± 0.104 | -0.227 ± 0.076 | 0.184 ± 0.105 | 0.376 ± 0.106 | -0.444 ± 0.116 | 1.12 ± 0.161 |
| H4 | 2.145 ± 0.09 | -1.25 ± 0.071 | 0.778 ± 0.101 | 1.899 ± 0.138 | -0.616 ± 0.041 | 5.4 ± 0.193 |
| I1 | 1.487 ± 0.11 | -0.377 ± 0.046 | 1.64 ± 0.093 | 2.438 ± 0.081 | -0.602 ± 0.031 | 7.01 ± 0.026 |
| I2 | 1.913 ± 0.026 | 0.108 ± 0.045 | 1.891 ± 0.024 | 1.143 ± 0.057 | 0.241 ± 0.038 | 6.88 ± 0.074 |
| I3 | 1.845 | -0.432 | 0.304 | 1 | -0.235 | 2.3 |

**Table S2: Expected prediction error under various experimental designs when the statistics are the full set of fitness values**

| Experimental design | Type of mutation | Landscapes using this protocol | n | | | $W_{max}$ | | | σ | | | Q | | |
|---|---|---|---|---|---|---|---|---|---|---|---|---|---|---|
| | | | rej | reg | nn | rej | reg | nn | rej | reg | nn | rej | reg | nn |
| 5 mutations, $2^5$ genotypes | random | A, C | 0.89 | 0.65 | 0.69 | 0.8 | 0.65 | 0.48 | 0.46 | 0.26 | **0.22** | 0.75 | 0.51 | 0.4 |
| | independently selected | - | 1.12 | 0.68 | 0.67 | 0.42 | 0.27 | **0.21** | 0.38 | 0.2 | **0.14** | 0.73 | 0.39 | **0.2** |
| | co-selected | F | 1.05 | 0.87 | 0.82 | 0.56 | 0.41 | 0.27 | 0.62 | 0.33 | 0.25 | 0.77 | 0.53 | 0.29 |
| 4 mutations, $2^4$ genotypes | random | - | 0.86 | 0.72 | 0.74 | 0.87 | 0.76 | 0.56 | 0.45 | 0.27 | 0.26 | 0.63 | 0.41 | 0.36 |
| | independently selected | I | 1.08 | 0.76 | 0.75 | 0.53 | 0.36 | 0.32 | 0.52 | 0.34 | 0.28 | 0.73 | 0.41 | 0.27 |
| | co-selected | G | 1.11 | 0.89 | 0.81 | 0.61 | 0.41 | 0.29 | 0.69 | 0.49 | 0.36 | 0.83 | 0.56 | 0.32 |
| 8 mutations, 8 single and 20 double mutants | random | - | 0.86 | 0.71 | 0.71 | 0.75 | 0.59 | 0.5 | 0.49 | 0.33 | 0.3 | 0.67 | 0.44 | 0.42 |
| | independently selected | - | 1.19 | 0.88 | 0.78 | 0.6 | 0.45 | 0.37 | 0.54 | 0.34 | 0.33 | 0.91 | 0.94 | 0.59 |
| | co-selected | - | 1.11 | 0.93 | 0.9 | 0.6 | 0.38 | 0.27 | 0.63 | 0.42 | 0.26 | 0.88 | 0.51 | 0.36 |
| 20 mutations, up to 121 genotypes | random | B | 0.75 | 0.89 | **0.62** | 0.73 | 0.59 | 0.38 | 0.45 | 0.36 | 0.3 | 0.75 | 0.54 | 0.49 |
| 9 mutations, 9 single mutants, 18 double mutants | independently selected | D | 1.09 | 0.75 | 0.76 | 0.58 | 0.36 | 0.24 | 0.5 | 0.27 | 0.25 | 0.74 | 0.75 | 0.48 |
| 6 mutations, 6 single mutants, 15 double mutants | independently selected | E | 1.15 | 0.71 | 0.83 | 0.6 | 0.36 | 0.28 | 0.63 | 0.39 | 0.39 | 0.8 | 0.76 | 0.54 |
| 5 mutations, $2^5$ genotypes | independently selected, high fitness combination | H1, H2 | 1.1 | 0.72 | **0.61** | 0.27 | 0.18 | **0.11** | 0.75 | 0.37 | 0.29 | 0.55 | 0.26 | **0.14** |
| 4 mutations, $2^4$ genotypes | independently selected, small fitness effect mutants | H3 | 1.05 | 0.7 | 0.71 | 0.74 | 0.61 | 0.53 | 0.57 | 0.39 | 0.33 | 0.88 | 0.51 | 0.48 |
| 4 mutations, $2^4$ genotypes | independently selected, large fitness effect mutants | H4 | 1.08 | 0.67 | **0.57** | 0.39 | 0.28 | **0.2** | 0.29 | 0.18 | **0.16** | 0.51 | 0.22 | **0.1** |

Prediction error for the four parameters of Fisher's model, for several experimental designs (based on single and double mutants, or complete sets of mutations and all associated genotypes) and selection procedures (random, independently selected, co-selected mutations), when the statistics used in the ABC algorithm are the full set of fitness values, and not summary statistics. For each parameter, the three lowest prediction errors are in bold, highlighting the protocol and inference algorithms that perform best.